\documentclass[structabstract,letterpaper]{aa}  

\usepackage{natbib}
\bibpunct{(}{)}{;}{a}{}{,}

\usepackage{graphicx}
\usepackage{txfonts}
\begin{document}
   \title{Very Long Baseline Array observations of the intraday variable source J1128+592}

%   \subtitle{}

   \author{K. \'E. Gab\'anyi
          \inst{1}
          \inst{2}
          %\inst{3}
          \and
          N. Marchili
          \inst{3}
          \and
          T. P. Krichbaum
          \inst{3}
          \and
          L. Fuhrmann
          \inst{3}
          \and
          P. M\"uller
          \inst{3}
          \and
          J. A. Zensus
          \inst{3}
          \and
          X. Liu,
          \inst{4}
          \and
          H.~G. Song 
          \inst{4}
          }

   \institute{Hungarian Academy of Sciences, Research Group for Physical Geodesy and Geodynamics, Budapest, Hungary \\ \email{gabanyik@sgo.fomi.hu}
              \and
              F\"OMI Satellite Geodetic Observatory, Budapest, Hungary
         \and
             Max-Planck-Institut f\"ur Radioastronomie, Auf dem H\"ugel 69, Bonn, Germany
          \and
          Urumqi Observatory, National Astronomical Observatories, Chinese Academy of Sciences, Urumqi 830011, PR China 
             }

   \date{}

% \abstract{}{}{}{}{} 
% 5 {} token are mandatory
 
  \abstract
{Short timescale flux density variations in flat spectrum radio sources are often explained by the scattering of radio waves in the turbulent ionized interstellar medium of the Milky Way.
One of the most convincing observational arguments in 
favor of this is the annual modulation of the variability timescale caused by the orbital motion of Earth around the Sun.
J1128+5925 is a recently discovered IDV source with a possible annual modulation in its variability timescale. New observations suggest a change in its variability characteristics.}
{We search for changes either in source structure or in the interstellar medium that can explain the variations in the IDV pattern of J1128+5925. Using  Very Long Baseline Interferometry (VLBI), we study a possible relation between source orientation on the sky and anisotropy angle seen in the annual modulation. Additionally, obtaining a source size estimate from VLBI data enables us to calculate the distance to the screen causing the variations in J1128+5925.}
{We observed the source in six consecutive epochs with the Very Long Baseline Array (VLBA) at three frequencies, 5\,GHz, 8\,GHz, and 15\,GHz in total intensity and polarization. This data are combined with our densely time-sampled flux-density monitoring performed with the radio telescopes at Effelsberg (Germany) and at Urumqi (China).}
  {The VLBA observations detected an east-west oriented core-jet structure with no significant motion in its jet. The expansion of the VLBI core leads to an estimate of mild relativistic speed ($2.5\textrm{\,c} \pm 1.4$\,c). The position angle of the VLBI jet agrees with the angle of anisotropy derived from the annual modulation model. No significant long-term structural changes were observed with VLBI on mas-scales, 
although, the VLBI core-size expansion offers a possible explanation of the observed decrease in the strength of IDV. VLBI polarimetry measured
significant changes in the electric vector position angle (EVPA) and rotation measure (RM) of the core and jet. Part of the  
observed RM variability could be attributed to a scattering screen ($37$\,pc distance) that covers the source (core and jet) and may be responsible for the IDV. 
Superposition of polarized subcomponents below the angular resolution limit may also affect the observed RM.
}
   {}


   \maketitle
%
%________________________________________________________________

\section{Introduction}

Intraday variability \citep[IDV, ][]{idv_discovery2, idv_discovery} is the rapid variation observed usually at cm wavelengths in flat spectrum quasars and blazars. The typical timescales of the variation range from less than an hour to a few days. Using causality and the common light-travel time argument, the variability timescales translate into source sizes on the $\mu$as scale and consequently into brightness temperatures in the range of $10^{15}-10^{21}$\,K. These values are far in excess of the inverse Compton-limit brightness temperature \citep{compton}. In this source-intrinsic interpretation,
relativistic boosting with large Doppler factors of $\gg 50$ would be required to reduce
these brightness temperatures to the inverse Compton limit of $10^{12}$\,K. However, such high values of Doppler factors were not observed in kinematic studies of quasars and blazars.

One solution to this controversy is the source-extrinsic theory of IDV. This explains that the rapid variations are caused during the propagation of radio waves. The variations are caused by interstellar scintillation (ISS) in the intervening turbulent, ionized interstellar medium of the Milky Way \citep[e.g.,][and references therein]{GBI}. In this scenario, the characteristic variability time-scale is inversely proportional to the relative velocity between the observer and the scattering medium. The so-called ``annual modulation'' of the variability 
timescale reflects the systematic variation in the relative velocity vector between orbiting Earth
and moving screen. The observation of systematic annual changes in the variability timescale
of an IDV source is regarded as strong evidence of scintillation induced variability
and a source extrinsic interpretation \citep{annual1819}. 
These seasonal cycles have been reported for several IDV sources: \object{J1819+3845} \citep{annual1819}, \object{0917+624} \citep{0917annual1, 0917annual2}, \object{PKS1519-273} \citep{annual1519_1, new_ann}, \object{PKS1257-326} \citep{bignall_newest}, and \object{PKS1622-253} \citep{new_ann}. 

Unfortunately, the detection of an annual cycle in IDV sources is not always straightforward:
in some IDV sources, a previous claim of an annual variability cycle could not be confirmed in subsequent year
\citep[][PKS0405-385]{0405annual}, and in one source at least ceased completely  \citep[][B0917+624]{cease_0917}. 
Since the ISS phenomena requires a source size of the order or smaller than the scattering size
of the screen, possible source intrinsic size variations
(e.g., expansion or blending effects of the scintillating component(s)) 
could explain these variations in the IDV pattern or even its disappearance 
\citep[e.g.,][]{cease_0917_2,cease_0917_3,cease_0917_4}. Alternatively, changes 
in the IDV characteristics may also be caused by changes in the scattering 
plasma \citep[e.g.,][]{0405annual}. One way of
distinguishing between these different possibilities, is to perform a polarimetric
VLBI monitoring at different frequencies.

J1128+5925 (hereafter J1128+592) is an intraday variable
source. Densely time-sampled flux density monitoring observations of the source have been carried out with the Effelsberg 100-meter (MPIfR, Germany) and the Urumqi 25-meter (China) radio telescopes 
at 5\,GHz over the past four years. In most of these observations, J1128+592 showed pronounced variability with peak-to-trough amplitudes
exceeding  20\,\%. The measured characteristic variability timescales were differ significantly from epoch to epoch, ranging from 0.2 to 1.6 days. 
The changes in the variability timescales are indicative of an annual cycle. 

After the discovery of IDV in J1128+592, \cite{utolag1} performed optical observation of the source. J1128+592 
did not show rapid variability in the optical bands. Additional observations \citep{utolag2} confirmed the optical quietness of the source.
The different variability behavior in optical and radio regimes is expected if the IDV in radio wavelengths caused by interstellar scintillation.

In the case of J1128+592, we observed in late 2007 and early 2008 a strong decrease in the IDV amplitudes. To find out 
whether source-intrinsic changes lead to a change in the variability behavior, 
we performed 6 epochs of Very Long Baseline Array (VLBA) observations between 2007 July and 2008 February.  
In this paper, we report on an analysis of these VLBA observations.

The paper is organized as follows. In Sect. \ref{sum}, we provide a brief summary of the single dish observations and the annual modulation model. In Sect. \ref{obs}, we describe the VLBA observations and the data reduction procedure. In Sect. \ref{tot}, we present the total intensity results of the VLBA observations, and in Sect. \ref{conn}, we discuss the IDV properties of J1128+592 in the light of the VLBA observations, and in Sect. \ref{pol} we present the polarized intensity results of the observations, which we discuss in Sect. \ref{poldisc}. Finally, in Sect. \ref{end} we summarize our findings.

The following cosmological parameters were used throughout this paper: 
$H_{0}=71\,\textrm{km\,s}^{-1}\textrm{Mpc}^{-1}\textrm{, } \Omega_{\textrm{matter}}=0.27 \textrm{, and } \Omega_{\textrm{vac}}=0.73$. 
At the the redshift ($z=1.795$) of the source, 1\,mas corresponds to a spatial
scale of $8.5$\,pc. An apparent angular separation rate of 0.1\,mas/yr corresponds to an apparent
velocity of $7.8$\,c.

\section{Summary of the single-dish observations and the annual modulation model \label{sum}}

Since 2004 December, J1128+592 has been observed 28 times at 5\,GHz
with either the Effelsberg 100-meter radio telescope (in 10 epochs) or
the Urumqi 25-meter radio telescope (in 18 epochs). In previous papers \citep{1128_AN, 1128_aa, 1128_kerastari},
parts of these observations were analyzed and discussed. Regarding the
observational techniques, data reduction, and timescale determination, we refer to those papers and references therein.

\begin{figure}
  \begin{minipage}[t]{\columnwidth}
    \begin{center}
      \includegraphics[width=\columnwidth]{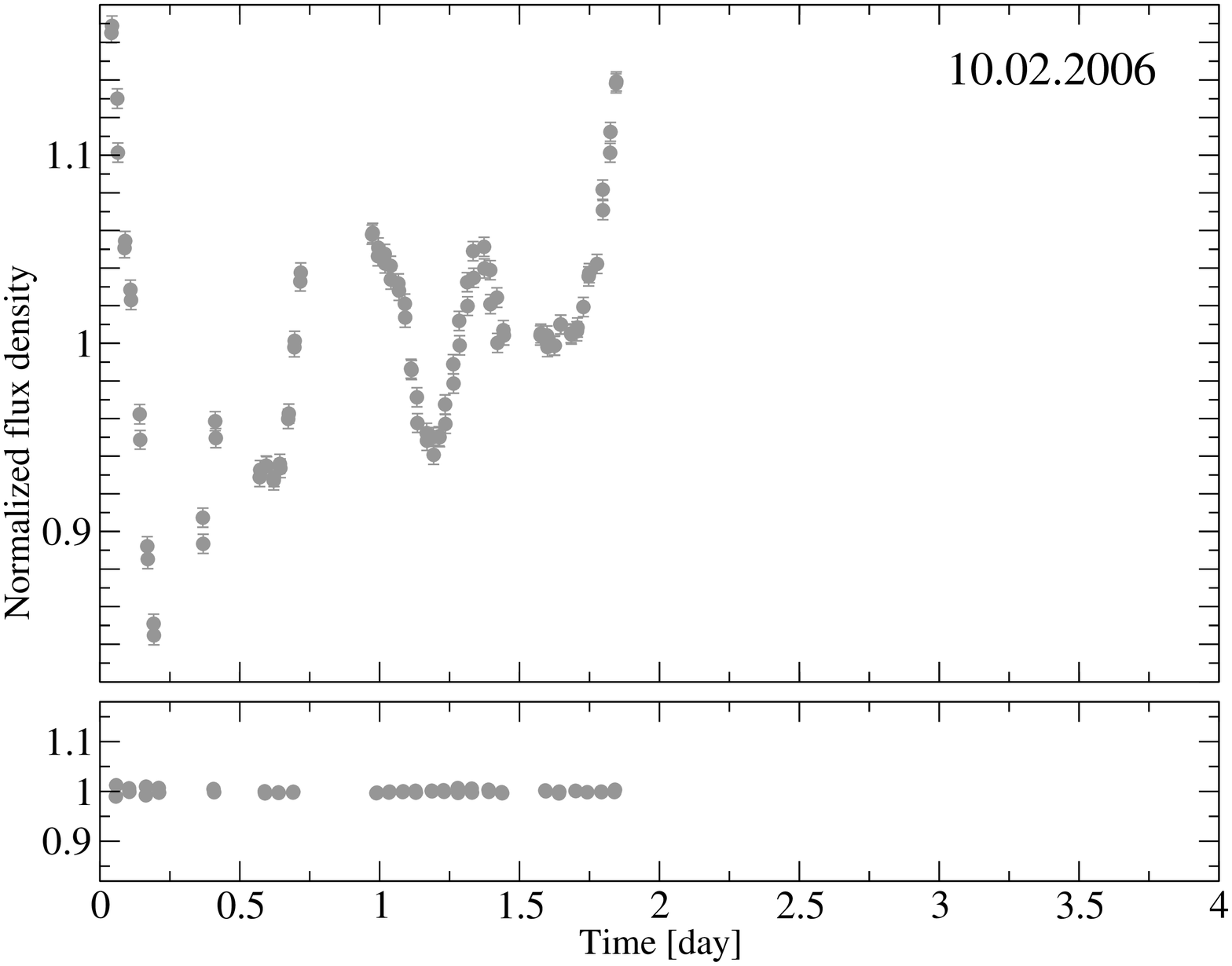}
    \end{center}    
  \end{minipage}
\hfill
  \begin{minipage}{\columnwidth}
    \begin{center}
      \includegraphics[width=\columnwidth]{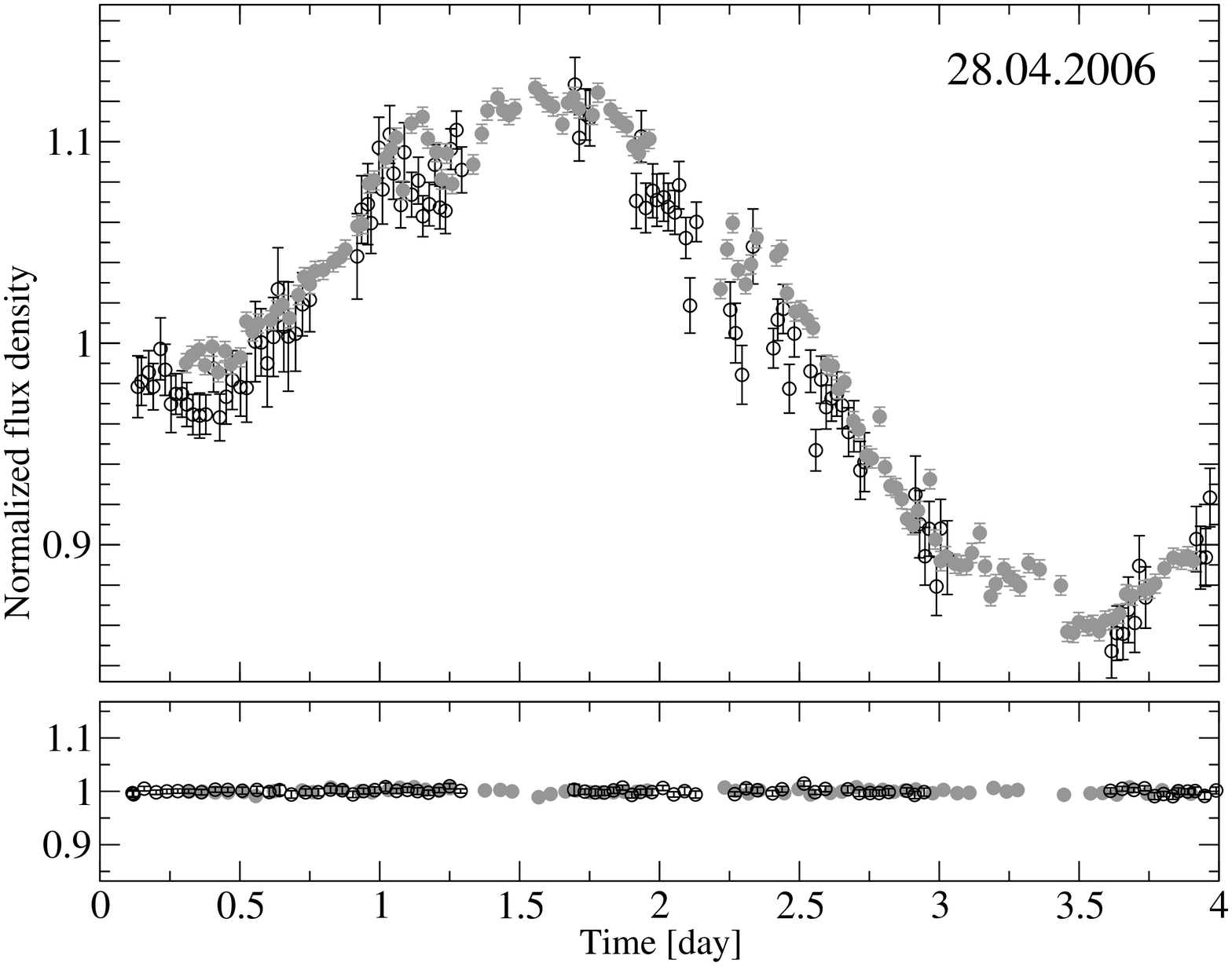}
    \end{center}    
  \end{minipage}
\caption{5\,GHz light curves of J1128+592, observed in February 2006 (top panel) and April 2006 (bottom panel) \citep{1128_aa}. Observations are performed with the Effelsberg antenna (gray symbols) and the Urumqi antenna (black symbols). In the lower parts of both panels, the light curve of a secondary calibrator source (B0836+710) is displayed for comparison.}
\label{fig:lcs}
\end{figure}

Two 5\,GHz light curves of J1128+592 are shown in Fig. \ref{fig:lcs}. The variability is pronounced, the peak-to-trough variability amplitude reaching 25\,\%. The two light curves exhibit different variability timescales of $\sim 0.2$\,day in February 2006 and more than $1.5$\,day in April 2006. We proposed that the changes in the characteristic variability timescale  are caused by annual modulation. In this case, the variability timescale is proportional to the scattering length-scale, and inversely proportional to the relative velocity between the scattering screen and the observer. 

However, the simple isotropic annual modulation model could not reproduce all observed changes in the variability timescale \citep{1128_aa}. In a more 
detailed %general 
scenario, the scintillation pattern could be 
more accurately described by an anisotropic model \citep{bignall_newest}. Here, the variability timescale also depends on both the ellipticity of the scintillation pattern and the direction in which the relative velocity vector ``cuts through'' the elliptical scintillation pattern (and not just the absolute value of the velocity vector).
We note that this model does not infer the origin of the anisotropy,
be it  source intrinsic or induced by the scattering medium.
A similar anisotropic model was used to explain the seasonal cycles in J1819+3845 \citep{annual1819}, PKS1257-326 \citep{bignall_newest}, and in both PKSB1519-273 and PKSB1622-253 \citep{new_ann}. We therefore fitted the anisotropic annual modulation model of \cite{bignall_newest} 
to our measurements with the following model parameters: 
the components of the screen velocity vector (in right ascension $\varv_\alpha$, and declination direction $\varv_\delta$), the scattering length scale ($s$, which depends on the screen distance and the scattering size), the position angle of anisotropy ($\beta$), and the axial ratio of the anisotropy ($r$).

\begin{figure}
\resizebox{\hsize}{!}{\includegraphics{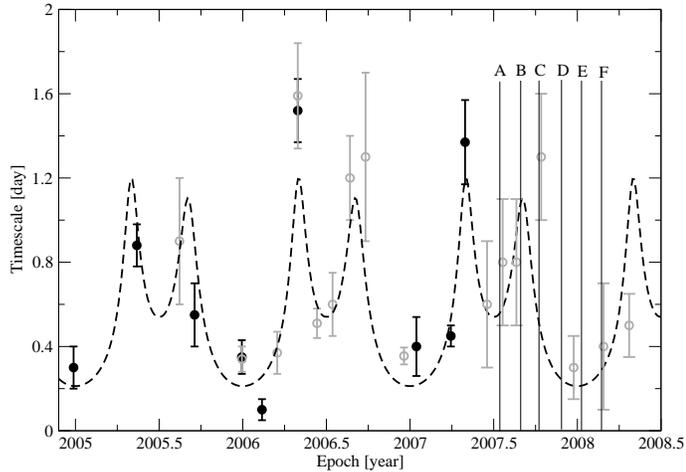}}
\caption{The variability timescales of the IDV of J1128+592, measured at different epochs. The dashed black line shows the 
fitted annual modulation model. Circles represent the data points used 
in the fitting. Open circles represent data from the Urumqi telescope, filled circles stand for Effelsberg measurements. 
Vertical lines mark the epochs of the VLBA observations.} \label{fig:ann_mod}
\end{figure}

As the monitoring observations continued and we gathered more data, we were able to refine the annual modulation model. For all available data acquired until 2008 April, we obtain the following parameters from the fits: $\varv_\alpha=(3 \pm 4) \textrm{\,km/s}$, $\varv_\delta=(-11 \pm 2) \textrm{\,km/s}$, $s=(1.3 \pm 0.3) \cdot 10^6$\,km, $r=3.4 \pm 0.8$, and $\beta=-98^\circ \pm 5^\circ$. The angle is measured north through east. The measured timescales and the fitted annual modulation model are displayed in Fig. \ref{fig:ann_mod}.

\begin{figure}
\resizebox{\hsize}{!}{\includegraphics{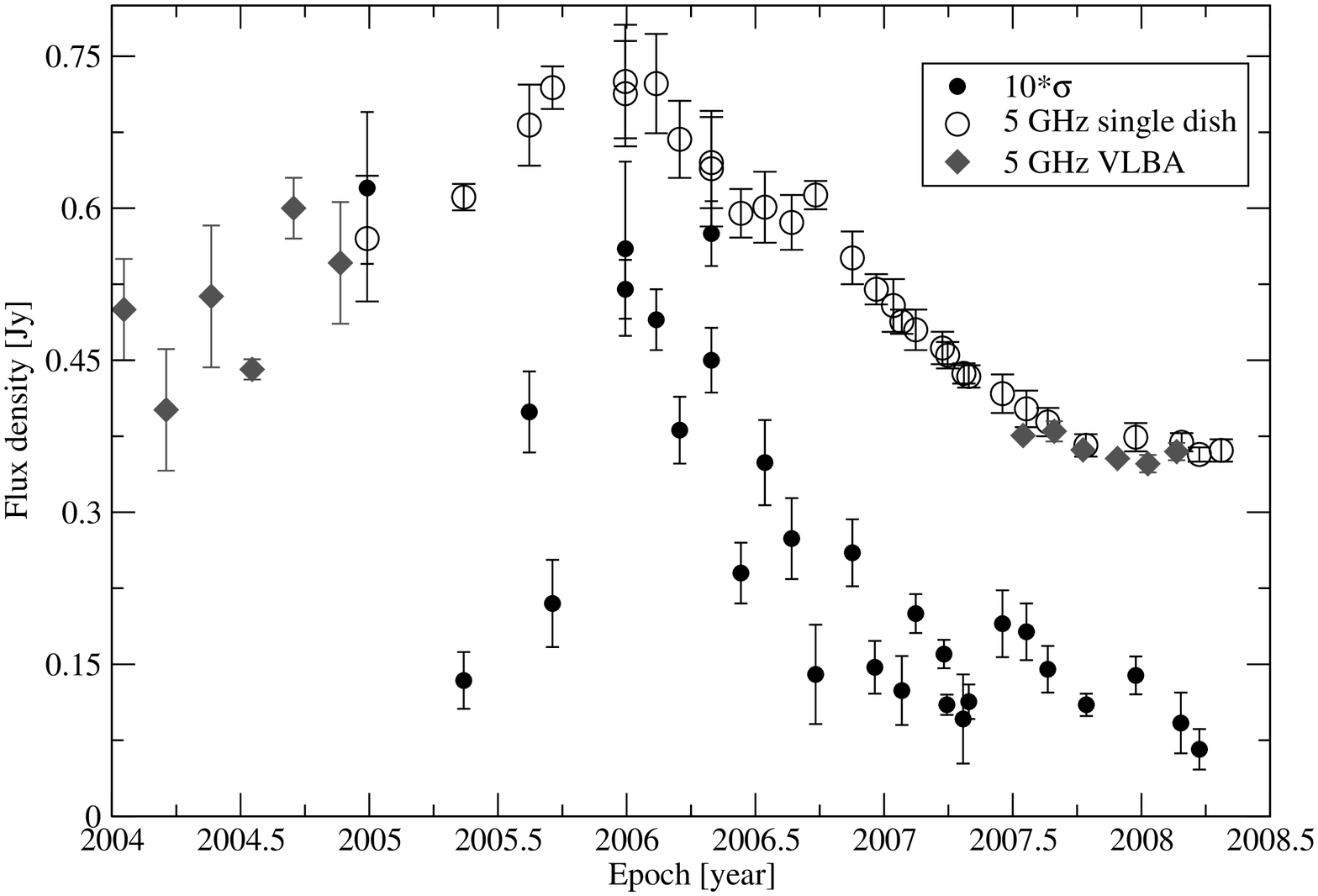}}
\caption{Flux density of J1128+592 as measured with single dish telescopes (open circles) and by the VLBA (filled gray diamonds) at 5\,GHz. Black dots represent the standard deviation of the flux density variations of J1128+592. The standard deviation values are multiplied by a factor of 10 
for a clearer display.}
\label{fig:sigma}
\end{figure}

During the 4 years of monitoring, the variability amplitude decreased. 
In Fig. \ref{fig:sigma}, we plot the total flux density at 5\,GHz versus time. We
also plot the standard deviation of the individual flux density measurements in a single observing
campaign, which is a measure of the ``strength of the IDV''. A value corresponding to ten times the 
standard deviation is displayed as filled circles in Fig. \ref{fig:sigma}.
We performed 6 epochs of VLBA observations during 2007 and 2008, to investigate the physical reasons behind the decreasing variability strength
and search for related structural variations in the source.

Comparing the average flux densities of these observations, we found that the source 
experienced significant long-term flux density change
(see Fig. \ref{fig:sigma}). The flux density of J1128+592 increased by $\sim 27$\,\% 
until the end of 2006, afterward the flux density continuously decreased until October 2007 by $\sim 63$\,\%. 
In parallel to the decreasing total flux density,
the strength of the variability (standard deviation given by filled circles in Fig. \ref{fig:sigma}) 
decreased in a similar way. We note that during the phase of brightening (before 2006),
J1128+592 was not monitored in a systematic way. Therefore, owing to differences
in the time sampling and duration of the individual measurements, the error bars and
the scatter in the measured standard deviation are larger for the first couple of measurements (for a discussion of the
long-term flux density variability, see Sect. \ref{tot}).

\section{VLBA observations and data reduction \label{obs}}

Our first VLBA observation took place in July 2007, and 5 subsequent epochs were followed by a 4 to 6 week separation. The last VLBA observation was
performed in February 2008. Each observation had a duration of 6 hrs and was performed at 5\,GHz, 8\,GHz, and 15\,GHz with dual polarization. 
The VLBI data were recorded with a 
data-rate of 256 Mbit/s. The sources \object{4C39.25} and \object{J1038+0512} were included as calibrators. In addition to the data
from our 6 VLBA experiments, we also searched the VLBA archive for previous observations of J1128+592. We (re)-analyzed six epochs 
of 5\,GHz observations closest int time to our monitoring program. J1128+592 was used as a calibrator in these archival observations. We unfortunately did not find 5\,GHz data overlapping with our single-dish campaign. The details of all these observations are summarized in Table \ref{tab:obs}. In the table, the first 6 epochs are 5\,GHz observations from the VLBA archive. The last 6 epochs are our multi-frequency observations.

\begin{table}
\caption{Details of the VLBA observations of J1128+592.}
\label{tab:obs}
\centering
\begin{tabular}{cccc}
\hline \hline
  & Epoch & $\nu$ [GHz] & Participating antennas \\
\hline
& 2004.047 & 5 & all \\
& 2004.211 & 5 & all \\
& 2004.386 & 5 & all \\
& 2004.545 & 5 & all \\
& 2004.706 & 5 & all except KP \\
& 2004.884 & 5 & all \\
\hline
A & 2007.540 & 5/8/15 & all, except HN at 15\,GHz \\
B & 2007.661 & 5/8/15 & all, except HN \\
C & 2007.773 & 5/8/15 & all, except SC \\
D & 2007.907 & 5/8/15 & all, except SC and PT \\
E & 2008.025 & 5/8/15 & all \\
F & 2008.137 & 5/8/15 & all \\
\hline
\end{tabular}
\end{table}

After correlation at the VLBA correlator in Socorro (NRAO), the data were calibrated and fringe-fitted using the standard procedures within the AIPS (Astronomical Image Processing System) software package. The post-processing of the data included the usual steps of editing, phase- and amplitude self-calibration, and imaging. These tasks were performed with the Caltech DIFMAP program package. The Caltech DIFMAP package was also used to fit the uv-data. We fitted the source structure in every epoch with circular Gaussian components. We tested fitting with elliptical Gaussians, however we did not obtain stable and physically meaningful results for every epoch (e.g., elliptical Gaussian components became degenerate). For consistency and to follow the same reduction and analyzing steps in every epoch, we parametrized the brightness distribution of J1128+592 by fitting only circular Gaussian components to the visibilities.

Calibration of the polarized data involved the correction for parallactic angle, determination of both right-left multiband and single-band delay corrections, feed D-term calibration, and the absolute calibration of the electric vector position angle (EVPA).
For D-term calibration, J1038+0512 was used. Its total intensity and polarization structure is simple and point-like, compared to our other calibrator, 4C39.25. The EVPA calibration was completed by comparing the averaged EVPA of 4C39.25 measured in the VLBA images and those values given by the Very Long Array (VLA) polarization monitoring program\footnote{{\tt http://www.vla.nrao.edu/astro/calib/polar}} at 5\,GHz and 8\,GHz. To calibrate EVPA at 15\,GHz, we used the 14.5\,GHz measurements of the University of Michigan Radio Astronomy Observatory \citep[UMRAO, e.g.,][]{umrao_ref}. The UMRAO observations are performed at three frequencies, 5\,GHz, 8\,GHz, and 14.5\,GHz.
The VLA and UMRAO results are all consistent, apart from one corresponding to the first epoch, for
which a difference of $\sim 100^\circ$ between the 8\,GHz EVPA of 4C39.25 was measured. 
The observations at the different sites were performed within two days. The measured polarized flux density differed significantly, the VLA flux density being twice as high as the UMRAO one. Since the amount of polarized flux density measured in our VLBA maps agreed with that measured for the UMRAO data, we decided to use the EVPA reported by the UMRAO.

\section{Results of the VLBA observations - total intensity \label{tot}}

\begin{figure*}
 \begin{minipage}[t]{0.33\textwidth}
 \begin{center}
  \includegraphics[width=5.0cm]{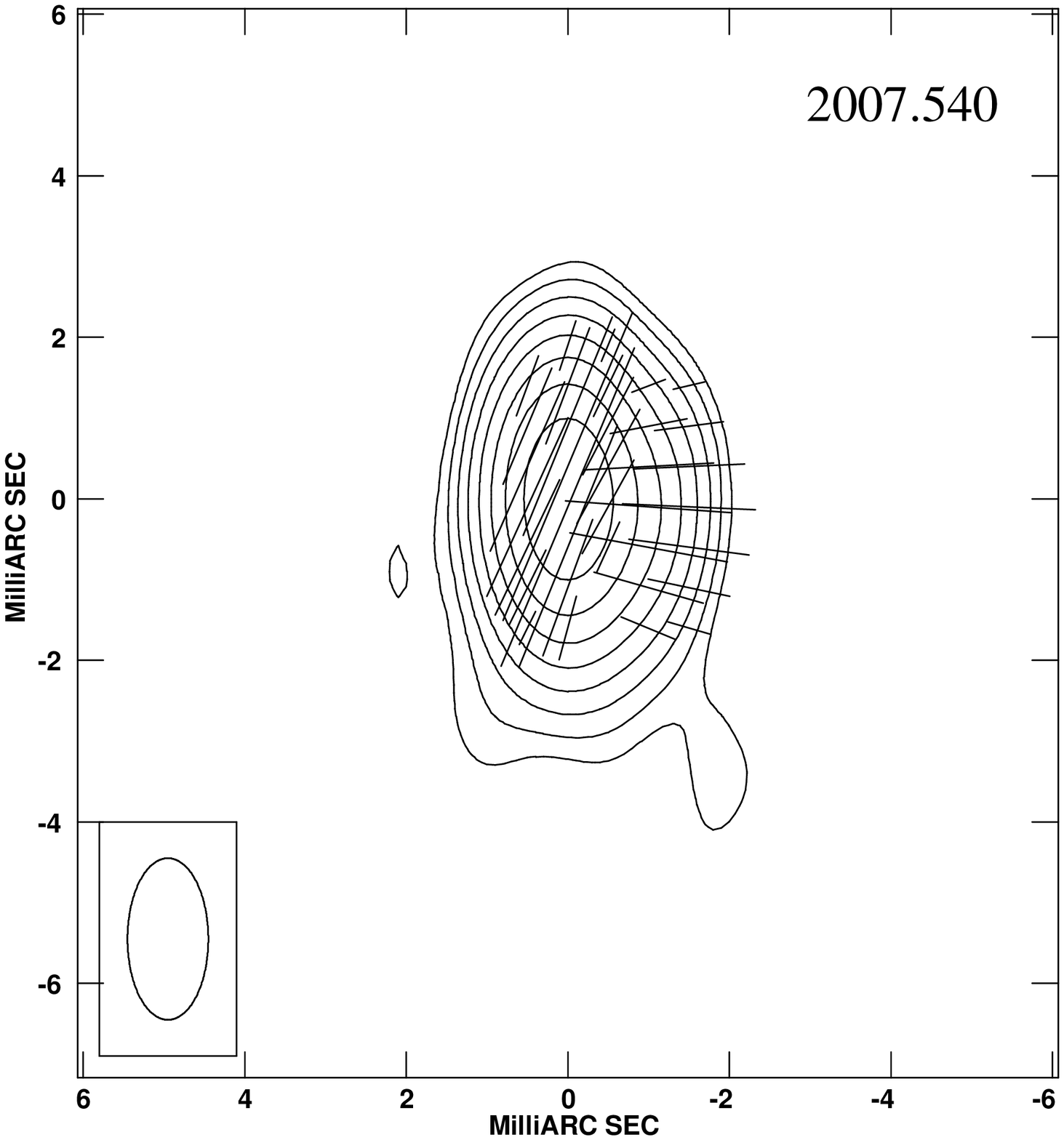}
 \end{center} 
 \end{minipage}
 \hfill
 \begin{minipage}[t]{0.33\textwidth}
 \begin{center}
  \includegraphics[width=5.0cm]{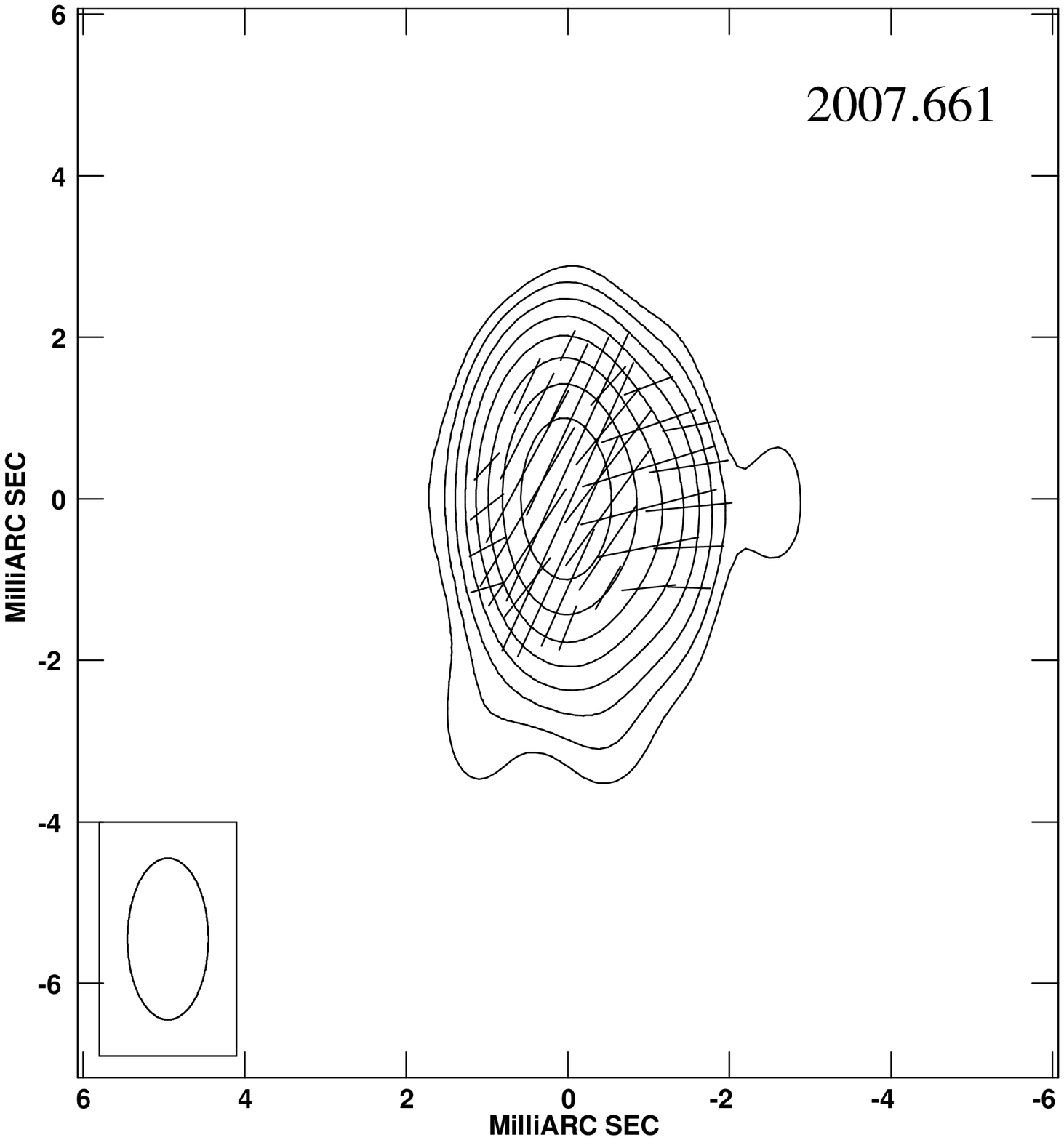}
 \end{center}
 \end{minipage}
\hfill
 \begin{minipage}[t]{0.33\textwidth}
 \begin{center}
  \includegraphics[width=5.0cm]{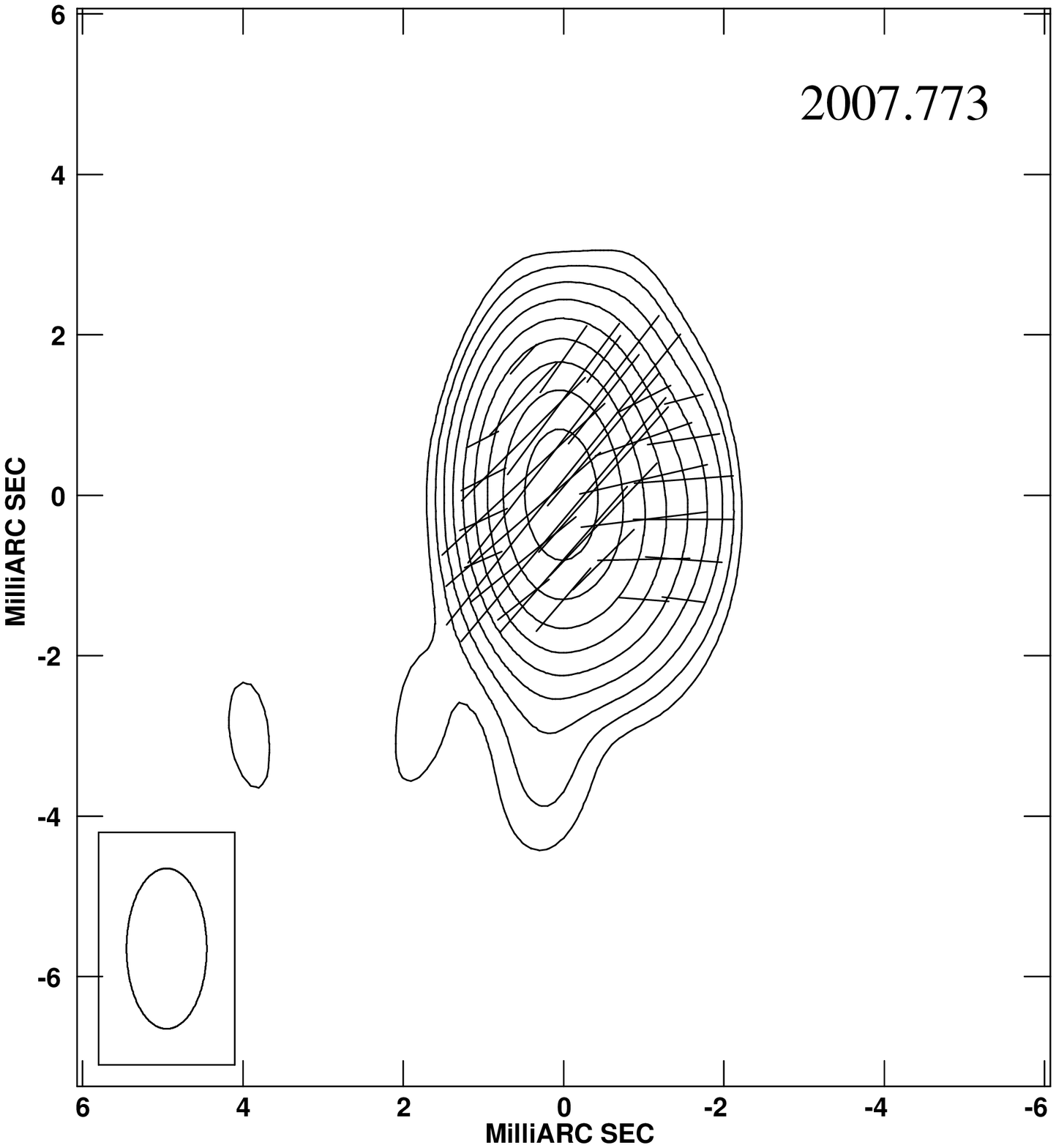}
 \end{center} 
 \end{minipage}
% \hfill
 \begin{minipage}{0.33\textwidth}
 \begin{center}
  \includegraphics[width=5.0cm]{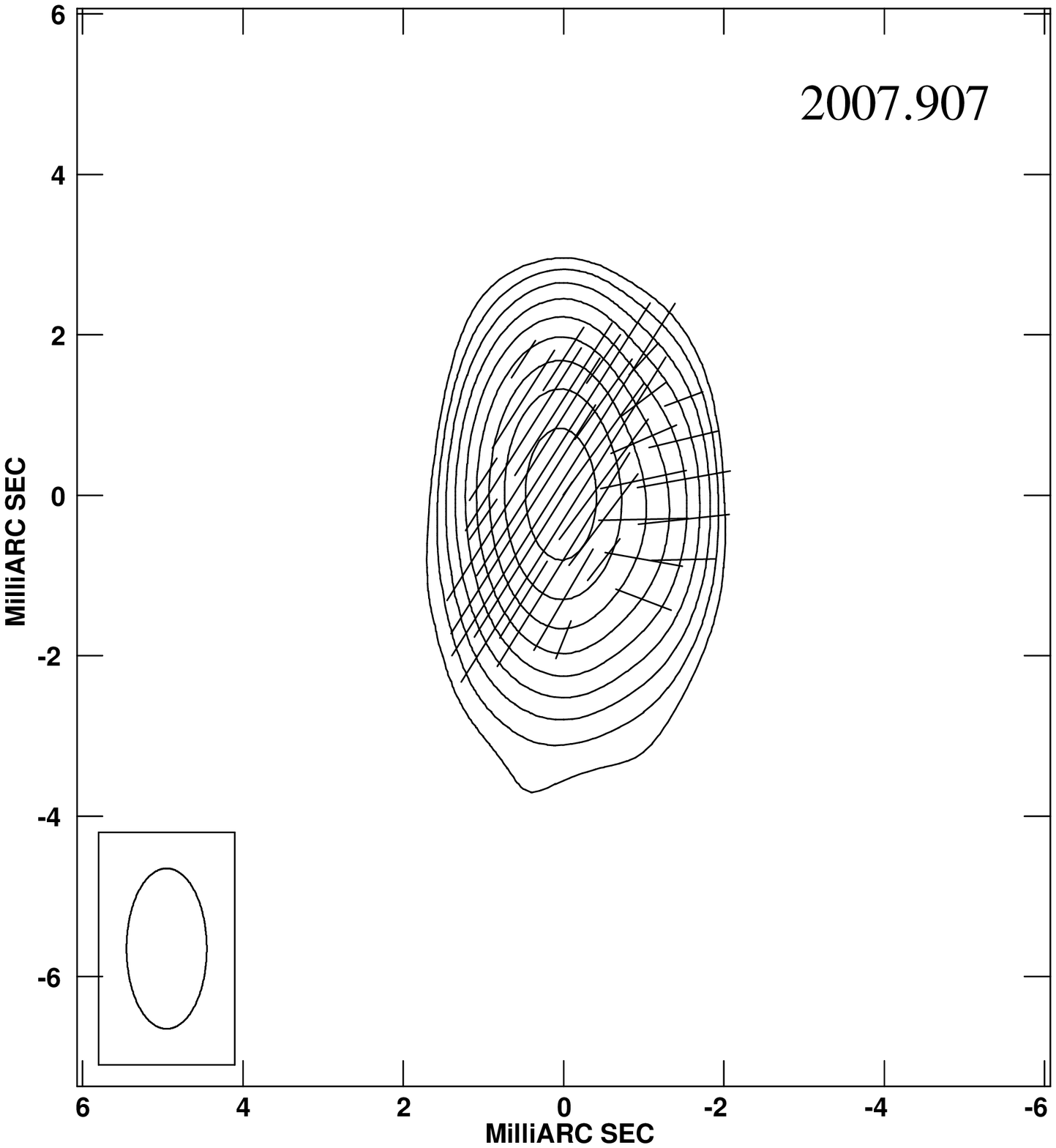}
 \end{center}
 \end{minipage}
%\hfill
 \begin{minipage}{0.33\textwidth}
 \begin{center}
  \includegraphics[width=5.0cm]{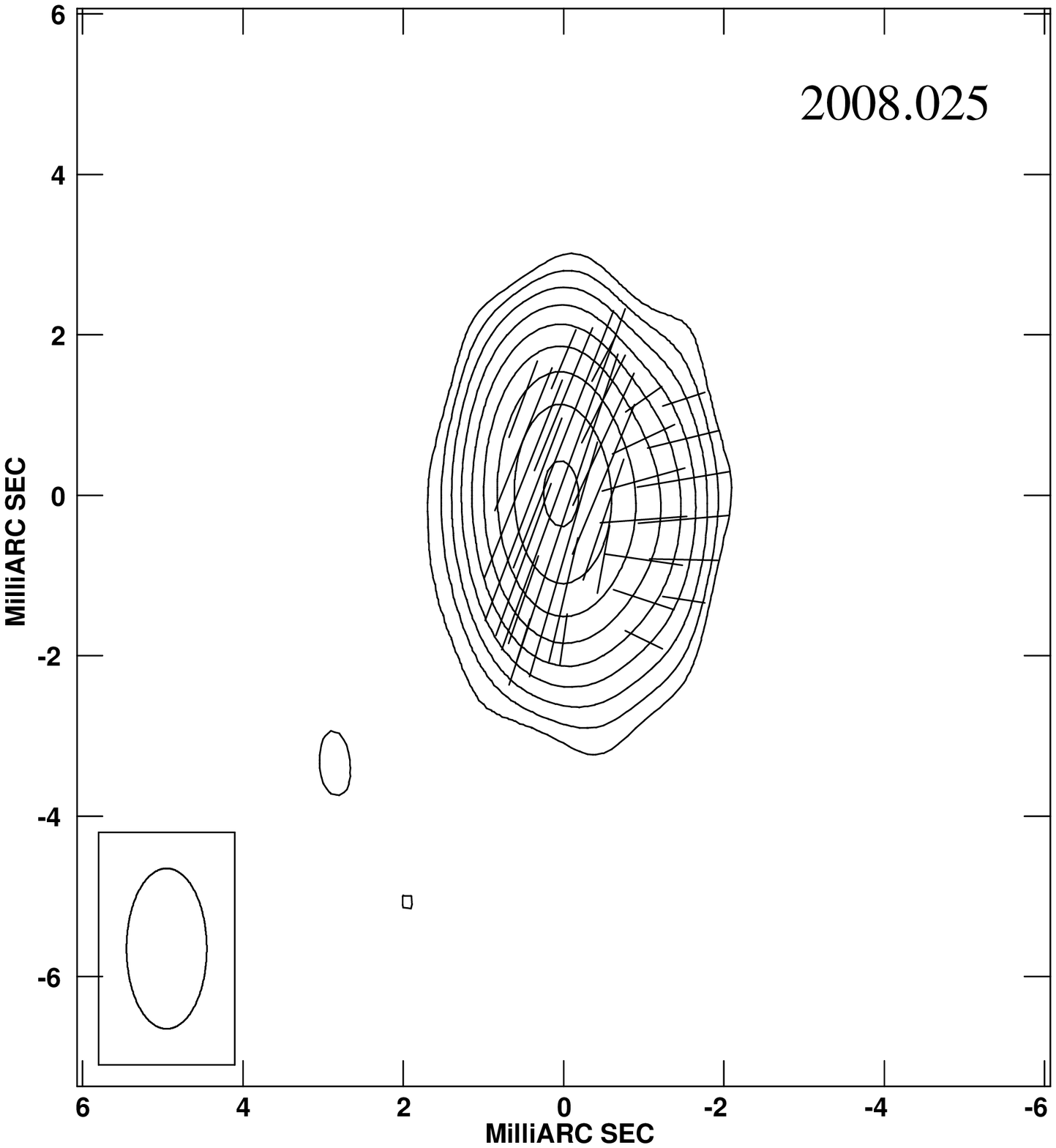}
 \end{center} 
 \end{minipage}
 \begin{minipage}{0.33\textwidth}
 \begin{center}
  \includegraphics[width=5.0cm]{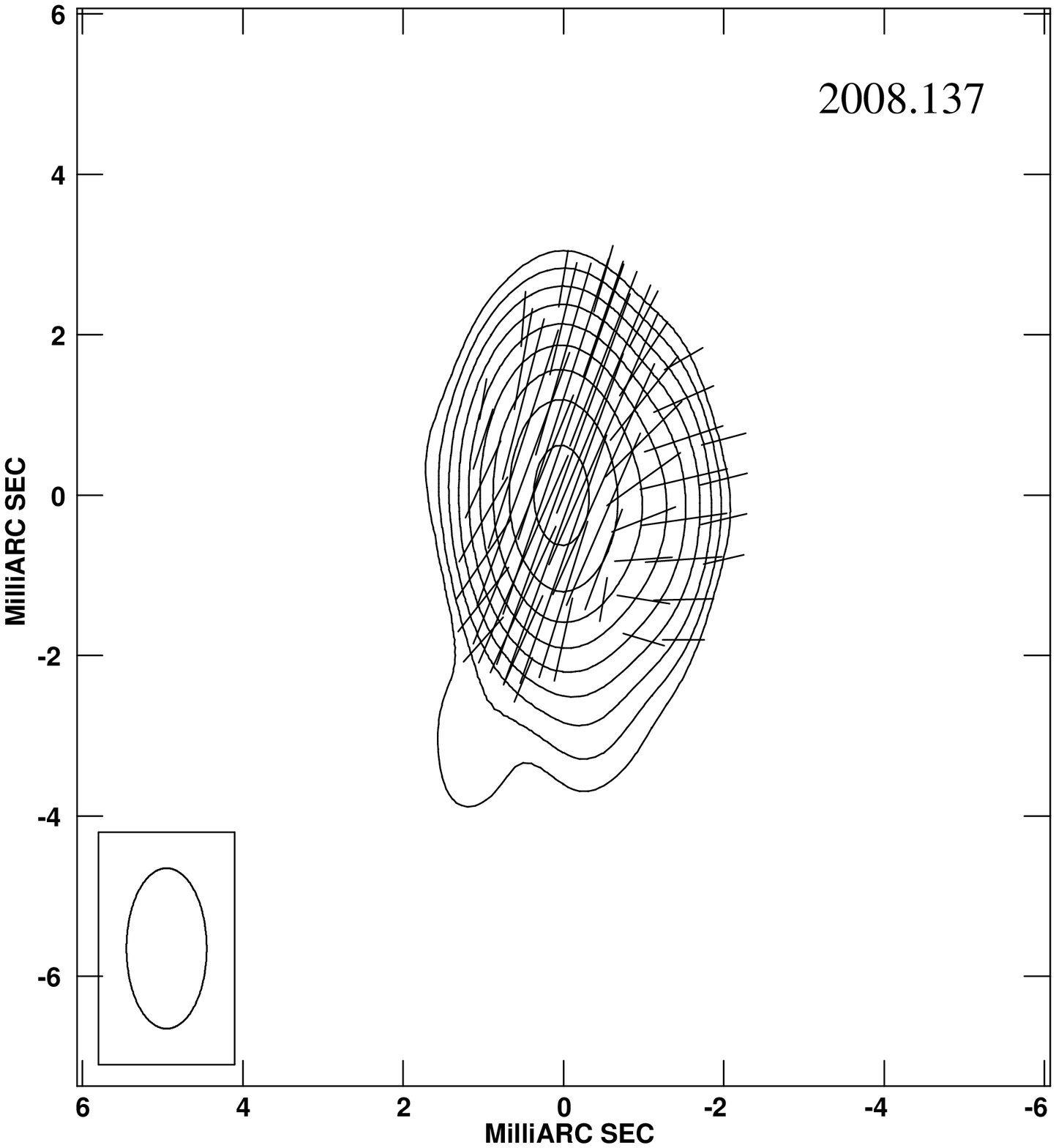}
 \end{center} 
 \end{minipage}
 \caption{5\,GHz VLBA maps of J1128+592 from the six observing epochs. Contours show total intensity, lines show the strength and direction of the polarized intensity. Contours are in percent of the peak flux and increase by factors of two.  The peak flux intensities are $312$, $299$, $295$, $292$, $277$, and $300$\,mJy/beam. The beam-size is 2\,mas x 1\,mas and is shown at the bottom left corner of each image. The 1 mas length of the superimposed polarization vectors corresponds to 1\,mJy/beam.} 
 \label{fig:5maps}
\end{figure*}

\begin{figure*}
 \begin{minipage}[t]{0.33\textwidth}
 \begin{center}
  \includegraphics[width=5.0cm]{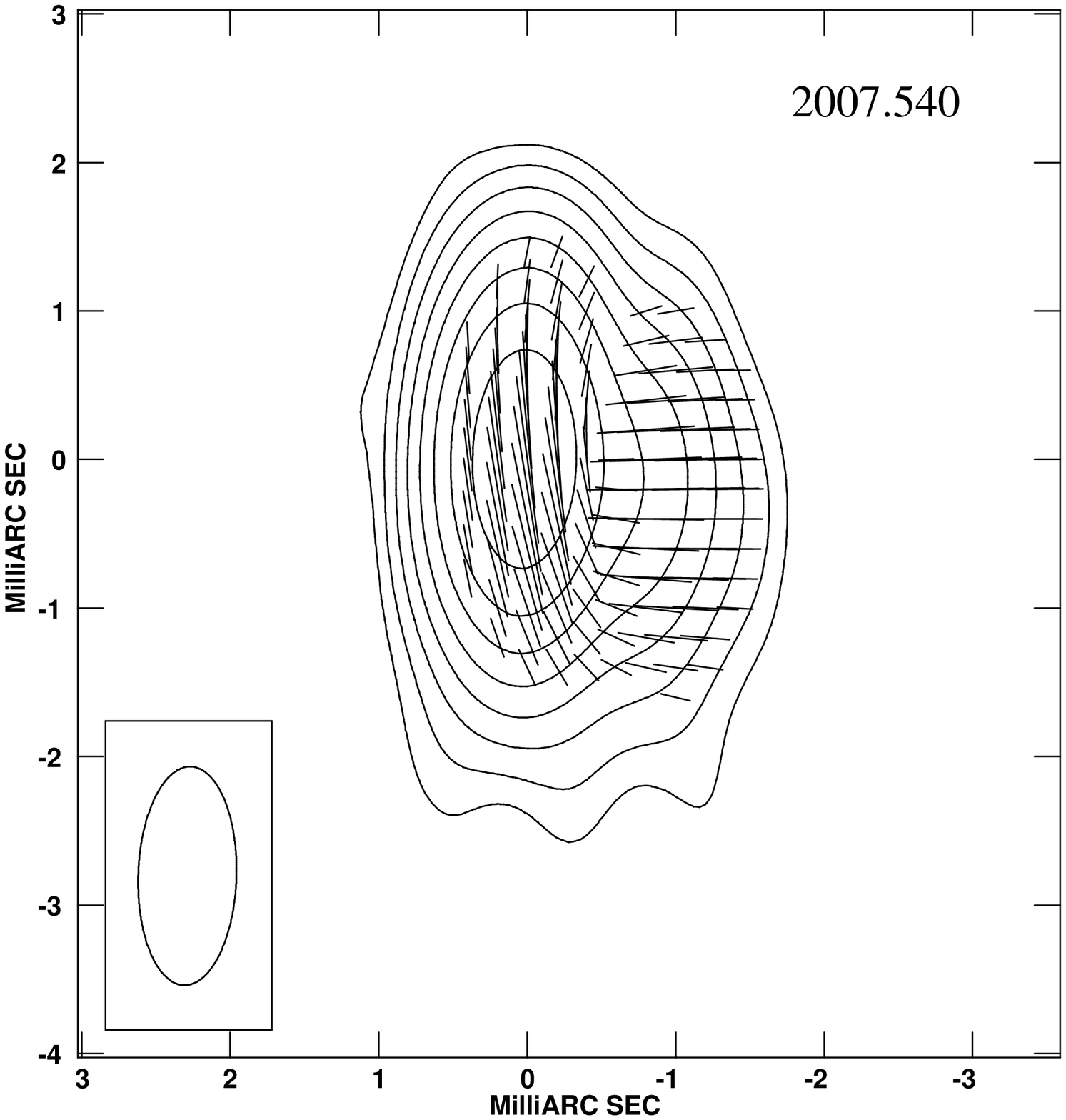}
 \end{center} 
 \end{minipage}
 \hfill
 \begin{minipage}[t]{0.33\textwidth}
 \begin{center}
  \includegraphics[width=5.0cm]{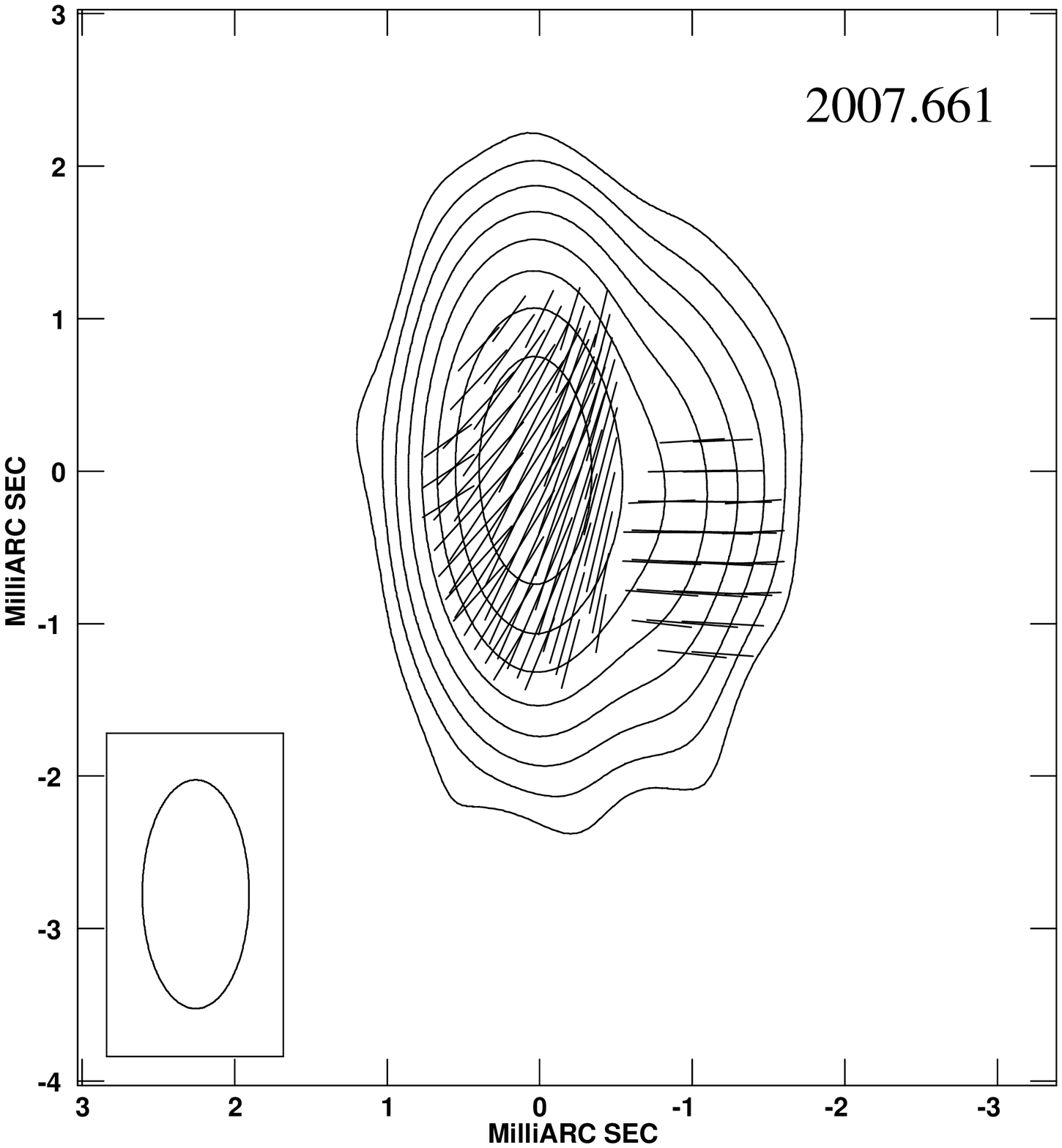}
 \end{center}
 \end{minipage}
\hfill
 \begin{minipage}[t]{0.33\textwidth}
 \begin{center}
  \includegraphics[width=5.0cm]{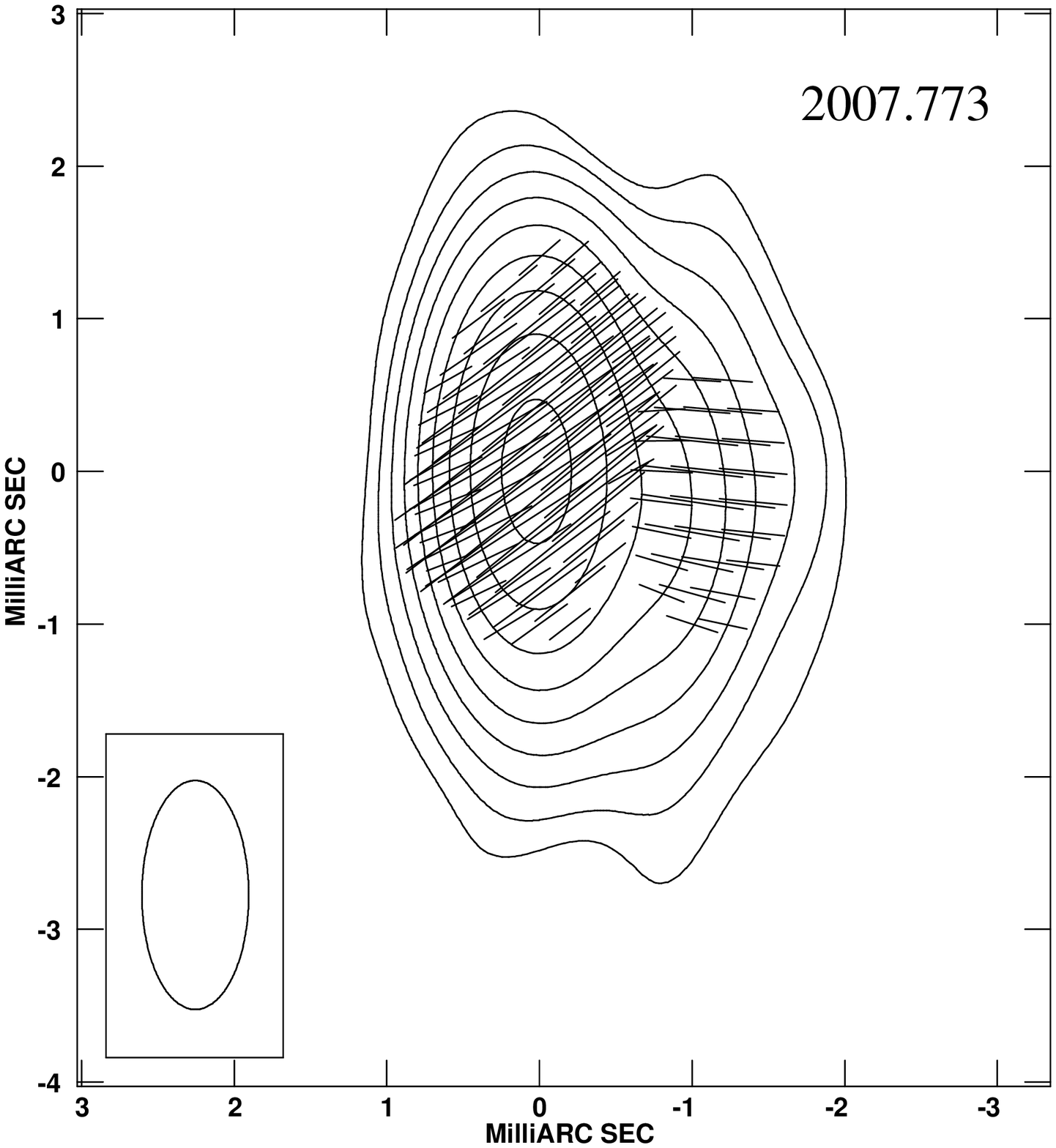}
 \end{center} 
 \end{minipage}
% \hfill
 \begin{minipage}{0.33\textwidth}
 \begin{center}
  \includegraphics[width=5.0cm]{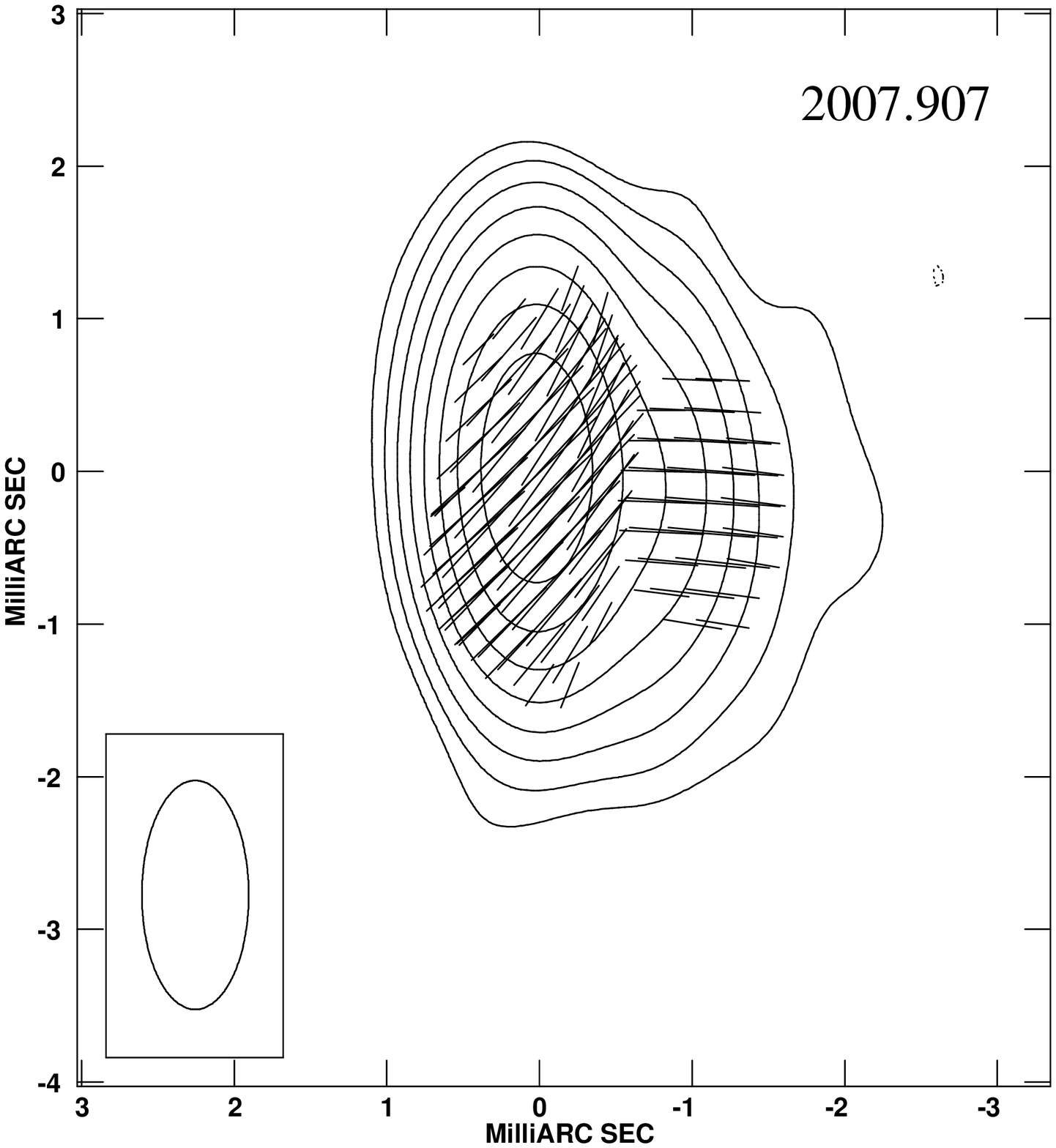}
 \end{center}
 \end{minipage}
%\hfill
 \begin{minipage}{0.33\textwidth}
 \begin{center}
  \includegraphics[width=5.0cm]{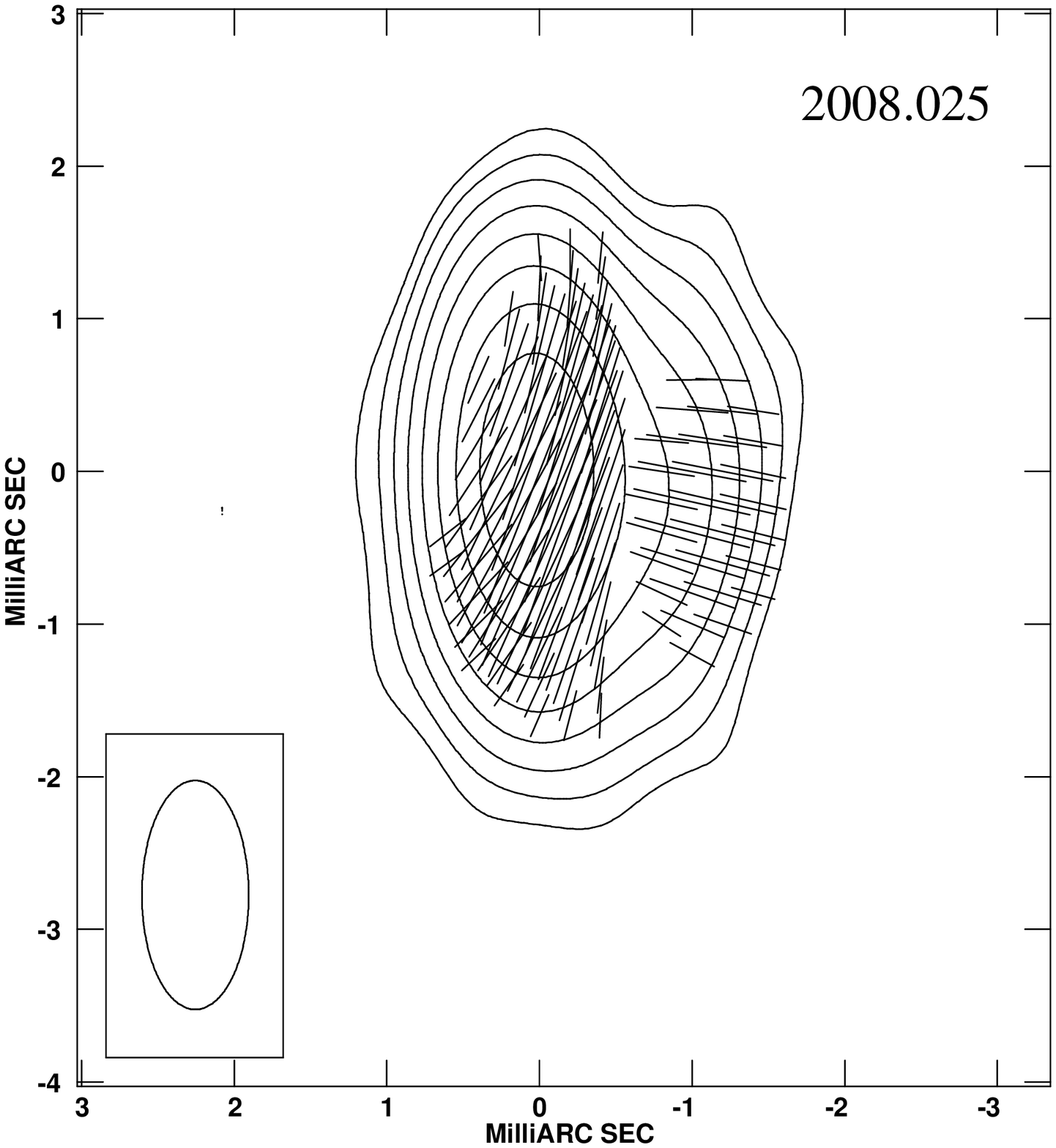}
 \end{center} 
 \end{minipage}
 \begin{minipage}{0.33\textwidth}
 \begin{center}
  \includegraphics[width=5.0cm]{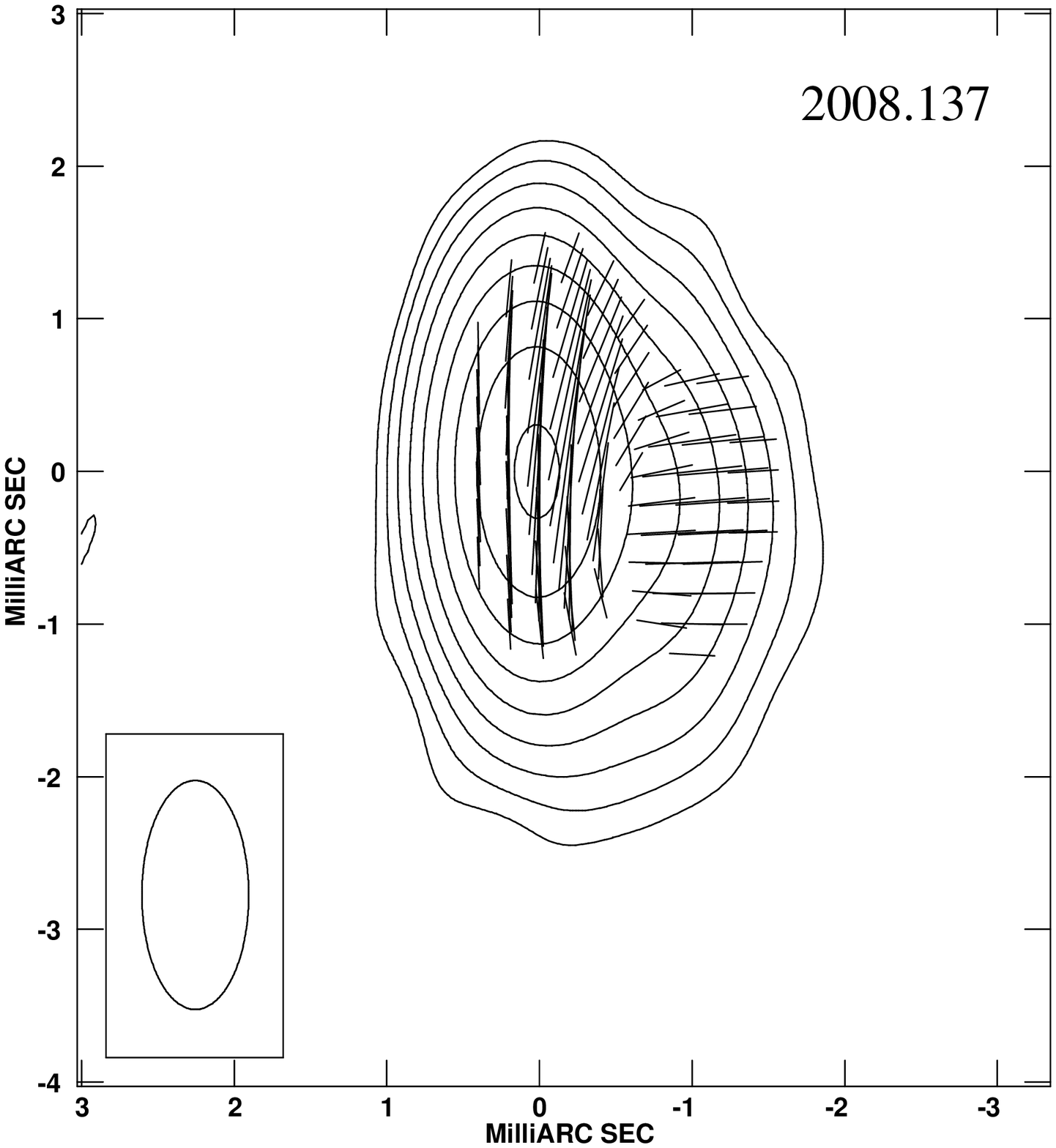}
 \end{center} 
 \end{minipage}
 \caption{8\,GHz VLBA maps of J1128+592 from the six observing epochs. Contours show total intensity, and lines indicate the strength and direction of the polarized intensity. Contours are in percent of the peak flux and increase by factors of two.  The peak flux intensities are $300$, $289$, $284$, $294$, $284$, and $283$\,mJy/beam. The beam-size is 1.5\,mas x 0.7\,mas and is shown in the bottom left-hand corner of each image. The 1 mas length of the superimposed polarization vectors corresponds to $2.5$\,mJy/beam.} 
 \label{fig:8maps}
\end{figure*}

\begin{figure*}
 \begin{minipage}[t]{0.33\textwidth}
 \begin{center}
  \includegraphics[width=5.0cm]{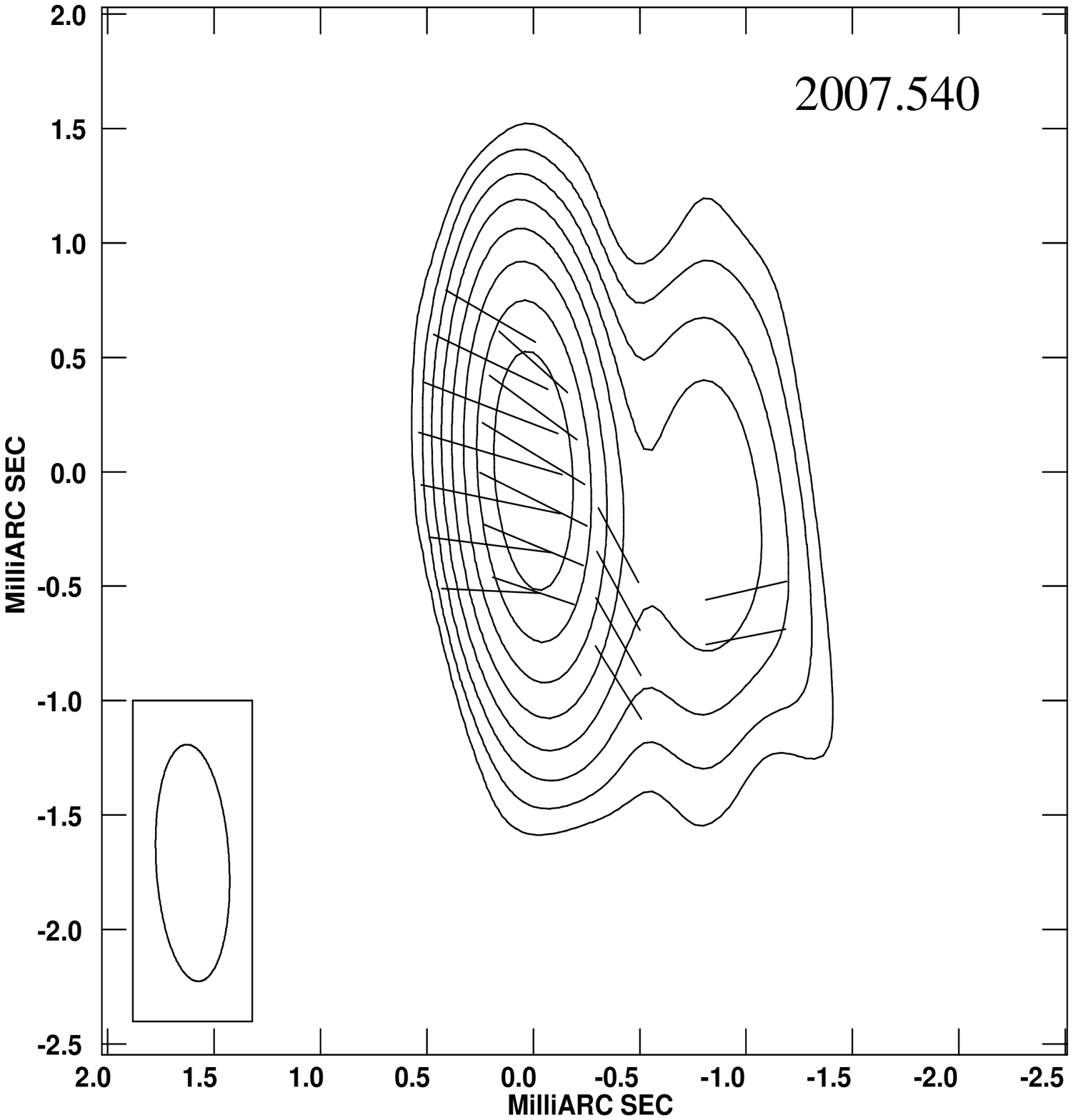}
 \end{center} 
 \end{minipage}
 \hfill
 \begin{minipage}[t]{0.33\textwidth}
 \begin{center}
  \includegraphics[width=5.0cm]{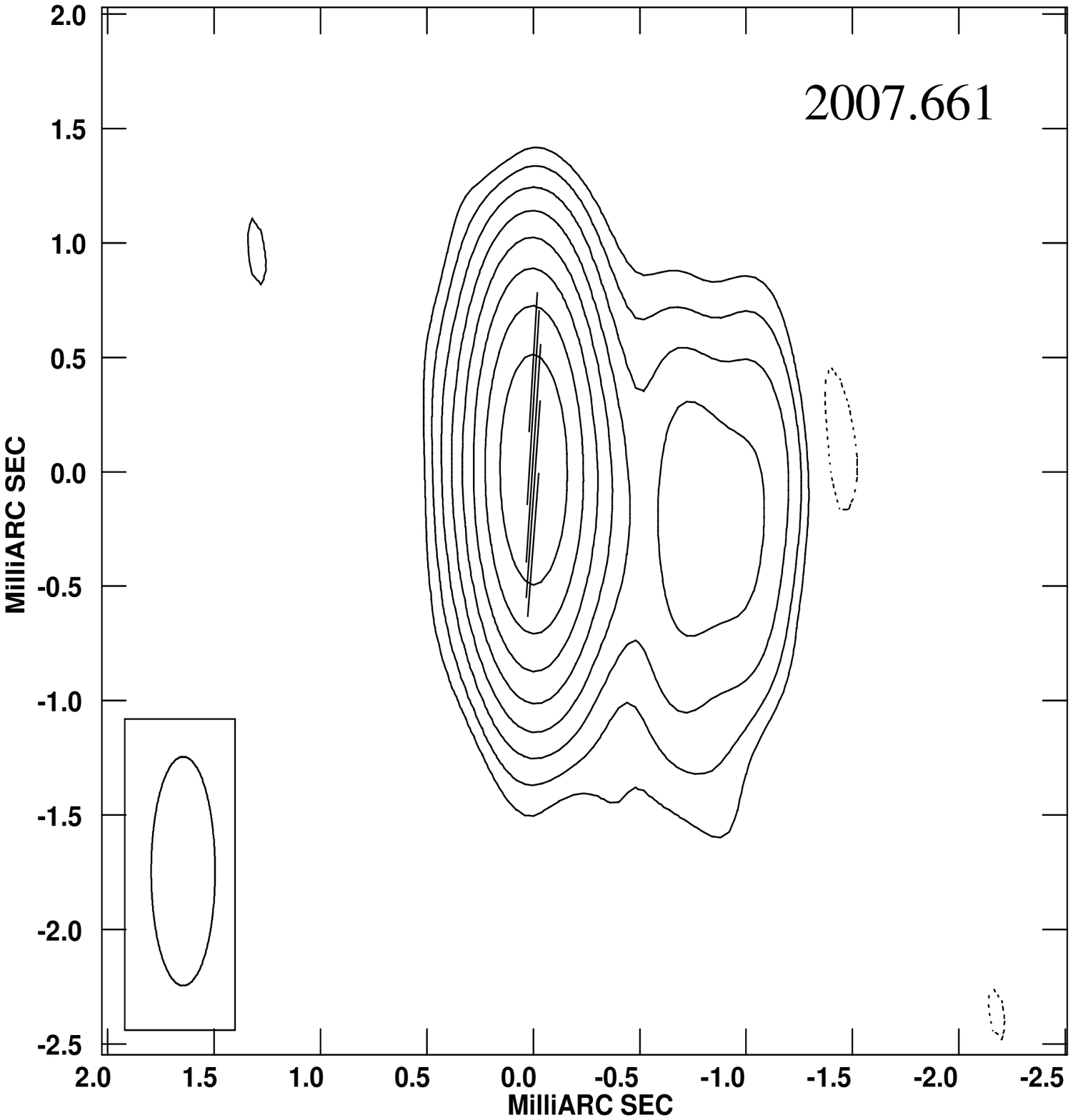}
 \end{center}
 \end{minipage}
\hfill
 \begin{minipage}[t]{0.33\textwidth}
 \begin{center}
  \includegraphics[width=5.0cm]{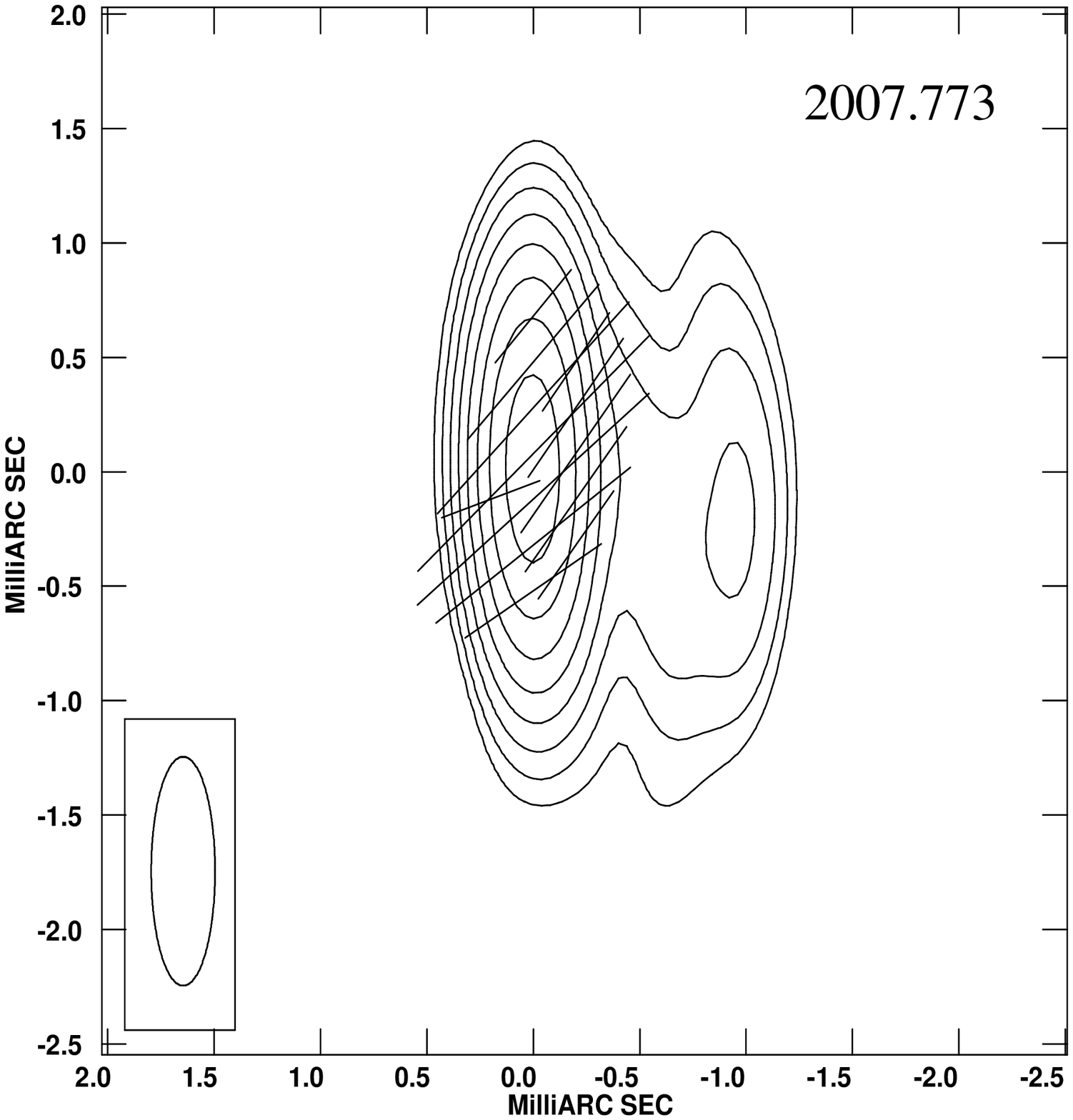}
 \end{center} 
 \end{minipage}
% \hfill
 \begin{minipage}{0.33\textwidth}
 \begin{center}
  \includegraphics[width=5.0cm]{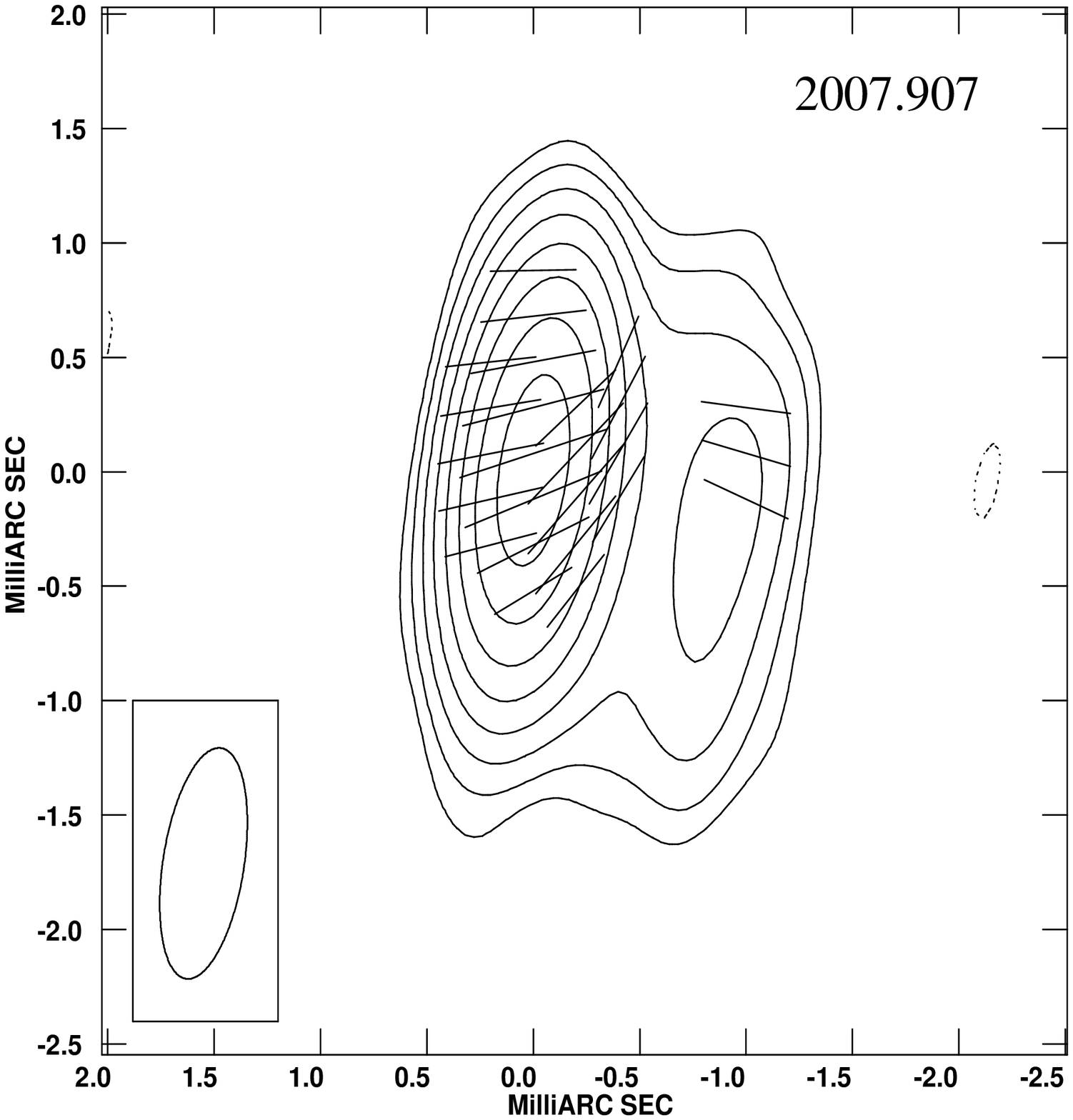}
 \end{center}
 \end{minipage}
%\hfill
 \begin{minipage}{0.33\textwidth}
 \begin{center}
  \includegraphics[width=5.0cm]{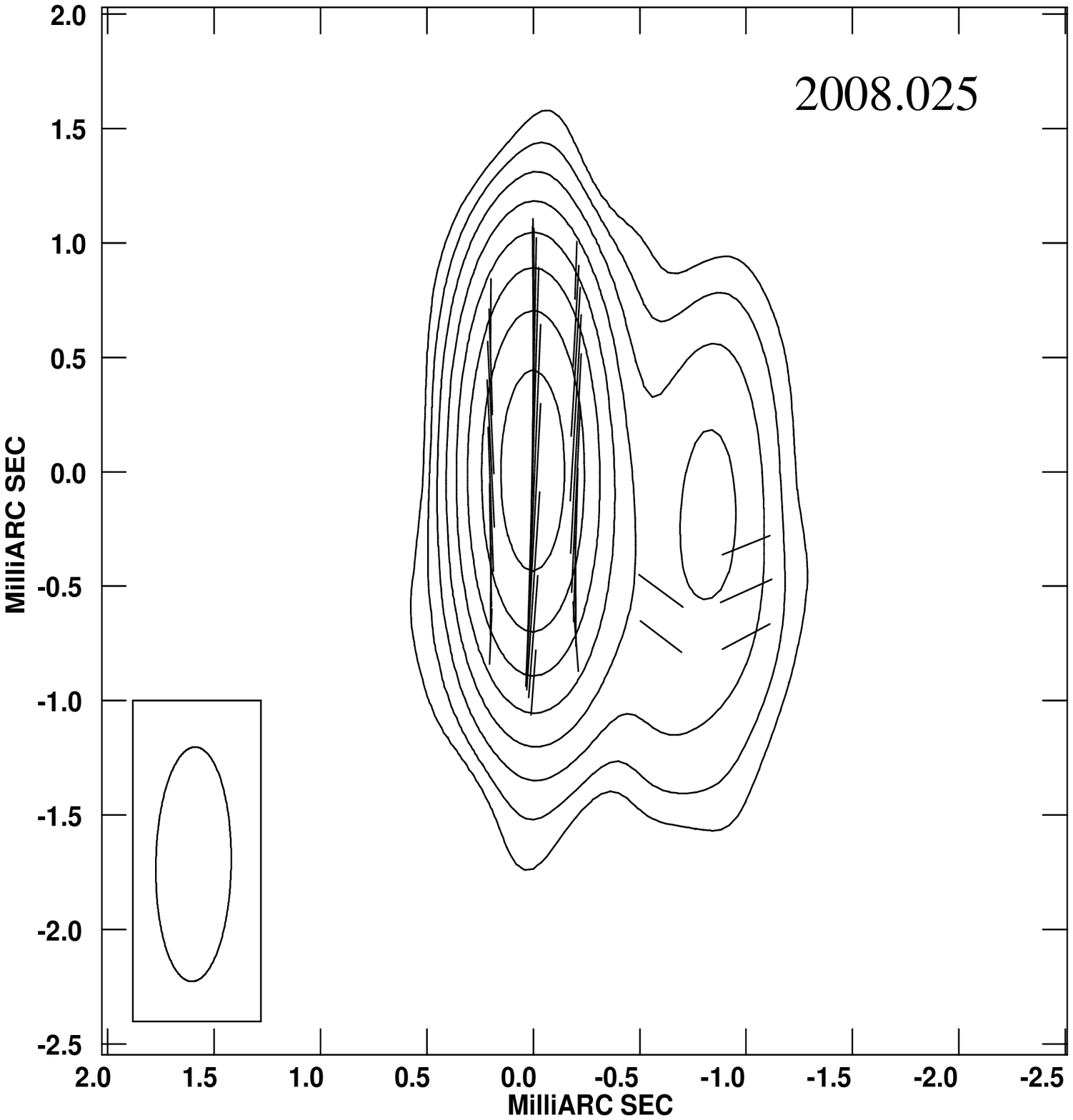}
 \end{center} 
 \end{minipage}
 \begin{minipage}{0.33\textwidth}
 \begin{center}
  \includegraphics[width=5.0cm]{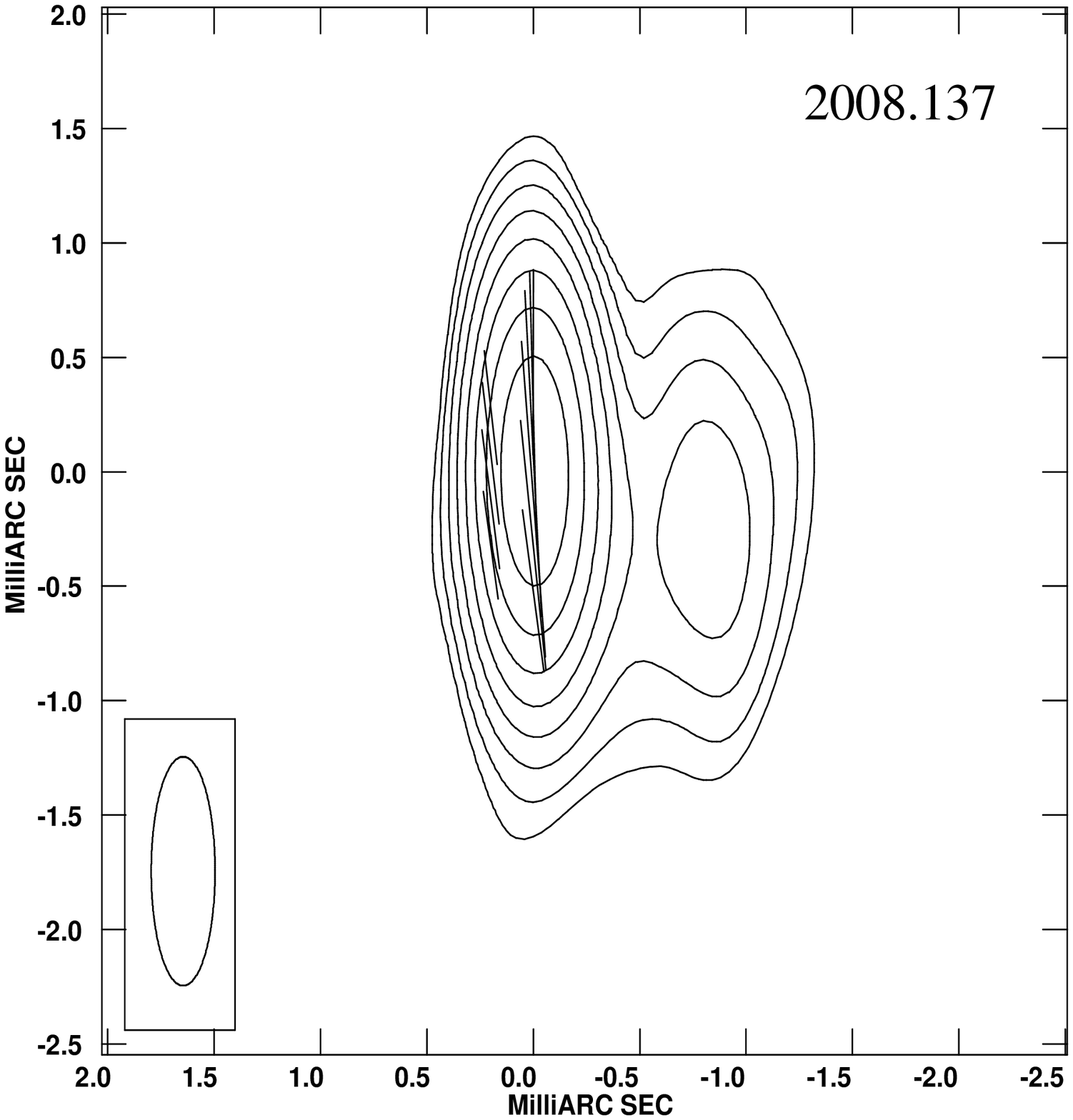}
 \end{center} 
 \end{minipage}
 \caption{15\,GHz VLBA maps of J1128+592 from the six observing epochs. Contours show total intensity, lines show the strength, and direction of the polarized intensity. Contours are in percent of the peak flux and increase by factors of two. The peak flux intensities are $232$, $229$, $220$, $236$, $232$, and $233$\,mJy/beam. The beam-size is 1\,mas x 0.3\,mas and is shown in the bottom left corner of each image. The 1 mas length of the superimposed polarization vectors corresponds to 2.5\,mJy/beam.} 
 \label{fig:15maps}
\end{figure*}

The aperture synthesis imaging of a non-stationary source, which varies during the
time of the observation, leads to a degradation in the reconstructed image. This 
amount of degradation depends on the variability amplitude and the timescale 
of the variability with respect to the total observing time. Simulations showed
that a smooth change in 80\,\% in the source flux density over an observing time of 12\,hrs
leads to residual side-lobes and a limitation of the dynamic range in the VLBI map
of the order of a few hundred \citep{var_VLBI}. In the case of J1128+592, the maximum
peak-to-trough variability amplitude during our 6 VLBA experiments was $\sim 16$\,\%
(Dec. 2007), but is typically lower and of the order of $\sim 10$\,\%. The variability
timescales were in the range 0.3 - 1.3\,days, which is extremely long compared to the 6\,hr 
duration of our VLBA experiments. We examined our VLBI images and we found no 
strong signs of image degradation (excessive rms, side-lobes, or symmetric structures indicative of
calibration errors). At 5\,GHz, where the IDV is strongest, we subdivided a VLBA
observation into two 3\,hr segments and imaged them separately (IDV decreases towards higher
frequencies, resulting in much lower image degradation at 15\,GHz). The resulting small
differences in the flux densities of the VLBI components ($\lesssim 10$\,mJy) are
within the errors introduced by the a-priori flux density calibration, which is
accurate at the $5-10$\,\% level, and the accuracy achieved by Gaussian model fitting,
which is mainly limited by the uv-coverage.
We therefore conclude that the IDV, which was present during our VLBI observations, did not affect 
the resultant images significantly, nor could it cause largely wrong estimates of the
VLBI component flux densities derived from Gaussian model fits. 

The VLBA maps (Fig. \ref{fig:5maps}, Fig. \ref{fig:8maps} and Fig. \ref{fig:15maps}) show the east-west oriented two-component 
structure of J1128+592, which is clearly resolved at 8\,GHz and 15\,GHz. 
At 5\,GHz, the total intensity maps show a point-like source at first glance, but after model-fitting with circular Gaussian components, a two-component model is found to provide a significantly better fit to the data in every epoch. 

The brighter and more compact component has a flat spectral index
between the 5\,GHz and 15\,GHz\footnote{The spectral index is defined as $S \sim \nu^\alpha$, where $\nu$ is the frequency and $S$ is the flux density.}. 
Its average spectral index during the
six epochs is $-0.06 \pm 0.04$. The western component has a steep
spectrum with an average spectral index of $-1.0 \pm 0.1$. Thus, we can conclude
that the brighter component is most probably the compact core, while the
westward feature is an optically thin jet component.

\begin{figure}
\resizebox{\hsize}{!}{\includegraphics{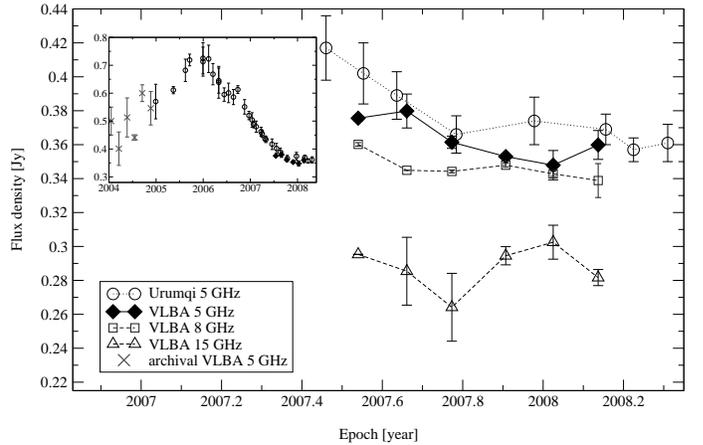}}
\caption{Flux density measurements of J1128+592 by single dish and VLBA. Circles represent the single dish flux density measurements. Diamonds, squares, and triangles show the summed model components of VLBA observations at 5\,GHz, 8\,GHz, and 15\,GHz respectively. In the insert, all of our 5\,GHz single dish observations (circles) are displayed together with our 5\,GHz VLBA data (diamonds), and archival 5\,GHz VLBA data (crosses).}
\label{fig:flux}
\end{figure}

The flux densities of the jet and core components at 5\,GHz, 8\,GHz, and 15\,GHz are listed in Tables \ref{tab:evpa5}, \ref{tab:evpa8}, 
and \ref{tab:evpa15}, respectively.
The summed flux density of the two fitted components at 5\,GHz agrees with the mean flux densities 
measured by the Urumqi telescope during that period (Fig. \ref{fig:sigma} and \ref{fig:flux}). More than 90\,\% of the flux density measured by a single dish is
recovered by our VLBA observations at 5\,GHz.

\begin{table}
  \caption{Flux densities of the fitted Gaussian model components in the six archival VLBA epochs.}
  \label{tab:arch_flux}
  \centering
  \begin{tabular}{c|cc}
    \hline
    \hline
    Epoch & $I_ \textrm{core}$ [mJy] & $I_\textrm{jet}$ [mJy] \\
    \hline
    2004.047 & $374\pm 25$ & $125 \pm 12$ \\
    2004.211 & $378 \pm 40$ & $23 \pm 2$ \\
    2004.386 & $398 \pm 80$ & $116 \pm 80$ \\
    2004.545 & $297 \pm 30$ & $144 \pm 14$ \\
    2004.706 & $546 \pm 35$ & $57 \pm 33$ \\
    2004.884 & $434 \pm 20$ & $112 \pm 10$ \\
    \hline
  \end{tabular}
\end{table}

The archival 5\,GHz VLBA data also shows an east-west oriented core-jet structure. However, the summed flux density of the model components are larger in every epoch than in our VLBA observations. This is in agreement with the long-term flux density changes revealed by the single-dish observations (Fig. \ref{fig:sigma}). According to the archival VLBA data (Table \ref{tab:arch_flux}), the flux density of the jet feature is very similar
to the results of our VLBA observations. But the core component is brighter than in our observations in 2007 and 2008, which
suggests that the core was responsible for the brightening observed in the single dish monitoring.

According to our VLBA observations, the size of the fitted circular core component is $\theta_\textrm{5\,GHz}=0.2 \pm 0.05$\,mas at 5\,GHz,
$\theta_\textrm{8\,GHz}=(0.08 \pm 0.02)$\,mas at 8\,GHz, and $\theta_\textrm{15 GHz}\sim 0.15$\,mas at 15\,GHz. In the standard jet model \citep[e.g.][]{jet_model}, the base of the jet is assumed to have a constant brightness temperature of $T_\textrm{B} \sim S/(\nu^2 \theta^2)  \sim \textrm{constant}$ (where $S$ is 
the flux density, $\nu$ the frequency, and $\theta$ is the FWHM source size). Thus, the size of a flat spectrum source can be expected to be inversely proportional to the frequency. However, the size of the flat spectrum core of J1128+592 does not follow this relation. The 15\,GHz size appears 
excessively large. This might indicate that the core component is a blend of the core and a new feature that is not (yet) resolved at 15\,GHz 
at the time of our observations. 

From the six reanalyzed archival epochs, we obtain an 
average core size of $\sim (0.07 \pm 0.02)$\,mas. This is less than half of the value that we measured in our VLBA observations. Unfortunately, we do not have VLBI data acquired simultaneously to the time when the source was in its most variable IDV state.
The available data, however, indicate that the core might have been more compact at that time. Since a larger size of the scintillating component(s) quenches the variations more \citep[e.g.,][and references therein]{acf_def2}, this increase may naturally explain the observed decrease in the variability 
amplitudes. The comparison with a standard scintillation model \citep{beckert_evn}
indeed suggests good agreement between a change of scintillating
source size (from $0.07$ mas to $0.2$ mas) and the observed change in variability amplitude (Fig. \ref{fig:sigma}).

From the model fits of the VLBA data, one can obtain an upper limit to the size of the VLBI core and its
flux density. Thus, one can calculate a lower limit to the brightness temperature of that feature. 
The redshift-corrected lower limit brightness temperature ($T_\textrm{B} \sim (1+z)S/(\nu^2\theta^2)$) of the core in our observations at 5\,GHz is $\geq 8.8 \times 10^{11}$\,K.
If the brightness temperature of the core is equal to this lower limit, assuming $10^{12}$\,K for the inverse-Compton limit
of the brightness temperature \citep{compton}, 
we find no need for relativistic beaming.
However, according to earlier VLBA data from the archive (year 2004), the redshift-corrected brightness temperature of the core at 5\,GHz 
was $\ge 11 \times 10^{12}$\,K at that time.  Here, at least a Doppler factor of $\sim 11$ is required to reduce the brightness temperature 
to the $10^{12}$\,K inverse-Compton limit.
This means that the minimum Lorentz-factor of the jet is $\gamma_{\rm min} =\delta_{\rm min}/2
\geq 5.5$.

To search for possible jet motion on mas-scales,
we checked for systematic changes in the relative separation of the two VLBI components of J1128+592.
At the highest available resolution, at 15\,GHz, we did not find any evidence of motion. A linear fit to the 15\,GHz data 
formally yields an angular separation rate of $0.03 \pm 0.08$\,mas/year. This translates into an upper limit
to the apparent speed of $(2\pm 6)$\,c.
At 15\,GHz, we observed
a slight change in the position angle of the jet feature relative to the core, which was assumed to be stationary: in the first three epochs, the 
position angle is $\sim -111^\circ $, while in the last three it is $\sim -107^\circ$. Compared to the average measurement
error in the position angle of $\sim 3^\circ$, we regard this change as insignificant.

\begin{figure}
\resizebox{\hsize}{!}{\includegraphics{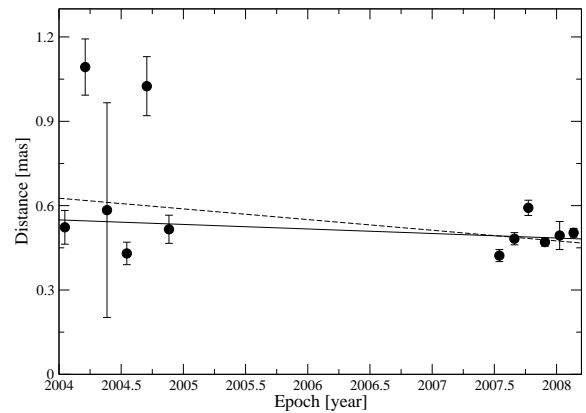}}
\caption{Relative separation of core and jet component of J1128+592 as measured with VLBI at 5\,GHz. The lines are fits from a linear regression 
analysis. The solid line represents a fit with linear weighting of the errors, the dashed line shows a fit with
quadratic weighting.}
\label{fig:distance}
\end{figure}

We also checked whether any significant motion can be detected between the 5\,GHz archival data from 2004 and our observations (2007/2008).
Resolving the VLBI structure of the two components in the archival data was more difficult 
because of the poorer uv-coverage (J1128+592 was a calibrator, usually observed in only a few scans.) 
Therefore, the derived distances (and position angles) between the components exhibited a larger scatter. 
A linear regression to the 5\,GHz data (weighting of the data by the inverse square of the errors) infers a formal speed 
of $(-0.02 \pm 0.02)$\,mas/year, which is consistent with either stationarity or marginal inward motion (Fig. \ref{fig:distance}).

Although no significant motion is seen in the jet, we can estimate the expansion velocity
from the observed change of the core size, and thus estimate the viewing
angle of the unresolved part of the jet relative to the observer. The core's FWHM changed by $0.13 \pm 0.07$ mas 
during four years, that is at a rate of
$0.03 \pm 0.02$\,mas/year. This translates into an apparent velocity of $(2.5 \pm 1.4)c$. 
The viewing angle is obtained from $\psi = \arctan{\frac{2 \beta_{app}}{\beta_{app}^2+\delta^2-1}}$.
Using the Doppler factor calculated from the brightness temperature of the core component ($11$),
we obtain for the viewing angle $\psi = 2.3^\circ \pm 1^\circ$.

In summary, we note that J1128+592 is at least a mildly ($\gamma \geq 5.5$) relativistic
source, which is oriented almost along the line of sight. It therefore shows a relatively
short jet (seen in projection) and a relatively large Doppler-factor of $\delta \geq 11$.

\subsection{Annual modulation model and VLBA source structure} \label{conn}

The orientation of the VLBI structure of J1128+592 is very similar to the orientation of the scattering ellipse 
deduced from the anisotropic annual modulation model. The average position angle of the 
jet feature is $\sim -116^\circ$ at 5\,GHz and $-108^\circ \pm 2^\circ$ at 15\,GHz (note, at 15\,GHz, that the position of the jet component
is located more accurately owing to a smaller beam size). While the position angle of
the scattering ellipse is $-98^\circ \pm 5^\circ$, according to the annual modulation model. 

Using the results of our VLBA observations, 
we fitted the variability timescales again, this time keeping the position angle of the anisotropy fixed at the position
angle of the jet feature ($-116^\circ$). The model parameters obtained agree well within the errors with 
the original unconstrained fit (see Sect. \ref{sum}). This may indicate that the anisotropy seen in the scintillation pattern 
originates mostly in the source structure rather than the scattering medium. 
However, additional observations are required
%would be needed 
to clarify unambiguously the roles of the scattering plasma and source structure in the
anisotropic scattering. According to \cite{acf_def}, a negative ``overshoot'' in the light-curve 
autocorrelations is indicative of anisotropy caused by the scattering medium. The study 
of the scintillation in all Stokes parameters ($I, U, Q, V$) can also shed light
on the origin of the anisotropy. This analysis however must await future 
single-dish observations of the polarized intensity of J1128+592, since the Urumqi telescope  
can only perform total intensity observations.

As mentioned in Sect. \ref{sum}, annual modulation models of different IDV sources always invoke anisotropic scattering. The 
ratio of anisotropy is usually above 4 \citep[e.g.,][report an axial ratio for PKS1257-326 $\sim 12$]{bignall_newest}. \cite{aniso} studied
whether an even higher degree of anisotropy can be used to describe the observations in the case of J1819+3845 and PKS1257-326.
The authors found that the anisotropy can be so large in these two sources  
that the scintillation pattern can be more effectively described one dimensional.
Therefore, they concluded it would be more likely that the anisotropy is caused by the scattering material 
rather than the source intrinsic structure.

We used the source model from the VLBI data (created by DIFMAP) to investigate how the velocity vector cuts 
through the source structure at the different observing epochs. We used an ``average'' source structure with a core-component flux 
density of $0.26$\,Jy, a core-component size of $0.2$\,mas, a jet-component flux density of $0.12$\,Jy, and 
a jet-component size of $1.1$\,mas. The position angle of the jet component was $-116^\circ$ and its distance from 
the core was $0.5$\,mas. In 2007.774 (VLBA epoch C), the source structure was aligned almost parallel to the velocity vector.
The direction of the velocity vector at that time of the year 
is $-126^\circ$ (see the ellipse in Fig. \ref{fig:vel}).
We cut through the source structure in this direction 
and fitted the resulting curves with a Gaussian. The resulting FWHM was $0.17$\,mas. 

Assuming this to be the value for the major axis of the scintillating source $\theta_\textrm{maj}$ and using a scintillation length-scale 
$s$ and axial ratio $r$ from the annual modulation model, we can calculate the screen distance $D$ by adopting the
equation $a_\textrm{maj}=s \cdot (2.54)^{-1}  \cdot \sqrt{r} = \theta_\textrm{maj} \cdot D$. The constant factor of $1/2.54$ is caused by
a different definitions of the variability timescale used by \cite{bignall_newest} and ourselves. More precisely, since the VLBA model fits
provide only an upper limit to the source size, the derived screen distance represents a lower limit. As
a lower limit to the distance of the scattering screen, we obtain $\sim 37$\,pc. 

This value is comparable to those derived for  
fast scintillators ($\le 10$\,pc), \citep{annual1819, bignall_newest}, is $3-4$ times lower than the
lower limits obtained previously for J1128+592 in \cite{1128_aa}. 
In  \cite{1128_aa}, we used a similar anisotropic annual modulation model, but we
estimated the source size from the frequency dependence of the variability strength at that time. 
This discrepancy can be explained by a not-so-precise estimate because of the
limited number of simultaneously measured multi-frequency light-curves and possible opacity effects.

\section{Results of the VLBA observations - polarized intensity \label{pol}}

\begin{figure}
  \resizebox{\hsize}{!}{\includegraphics{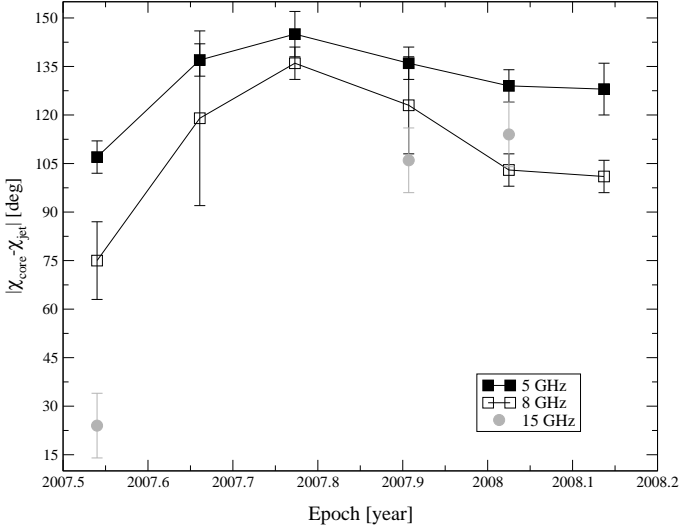}}
\caption{The difference between the core and jet EVPA at all three observing frequencies measured for the six VLBA epochs. Filled black squares represent 5\,GHz data, open black squares represent 8\,GHz data, and gray dots represent 15\,GHz data.}
\label{fig:polpa}
\end{figure}

Both the core and the jet component are polarized at 5\,GHz and 8\,GHz. The polarization of the jet component at 15\,GHz can be detected 
above the noise level in only three epochs (in 2007.54, 2007.907, and 2008.025).

The peak of the total intensity and the polarized feature in the core coincides in every epoch and frequency, except in 2007.54 at 15\,GHz. 
In this data set, the polarized intensity feature is displaced by $\sim 0.15$\,mas to the east from the total intensity peak.

The polarized flux densities and the EVPAs of the core and the jet are summarized in 
Tables \ref{tab:evpa5}, \ref{tab:evpa8}, and \ref{tab:evpa15}.
At 15\,GHz, we show values corresponding to the polarized patches at the largest core separation.

\begin{table*}
\begin{minipage}{\linewidth}
\caption{The results of the 5\,GHz VLBA observations.}\footnotetext{Col. 1 lists the observing epochs; Col. 2 lists the integrated flux density of the core component obtained from DIFMAP model-fitting; Col. 3 lists the polarized flux density of the core component; Col. 4 lists the EVPA of the core; Col. 5 lists integrated flux density of the jet component obtained from DIFMAP model-fitting; Col. 6 lists the polarized flux density of the jet component; Col. 7 lists the EVPA of the jet; and Col. 8 lists  the difference between the EVPA of the core and the jet.}
\label{tab:evpa5}
\centering
\begin{tabular}{c|ccc|ccc|c}
\hline
\hline
 Epoch & \multicolumn{3}{c}{Core} & \multicolumn{3}{c}{Jet} & \\
 & $I_\textrm{c}$ [mJy] & $P_\textrm{c}$ [mJy] & $\chi_\textrm{c} [^\circ]$ & $I_\textrm{j}$ [mJy] & $P_\textrm{j}$ [mJy] & $\chi_\textrm{j} [^\circ]$ & $\vert \chi_\textrm{c}-\chi_\textrm{j} \vert [^\circ]$ \\
\hline
2007.540 & $286 \pm 5$ & $3.2 \pm 0.3$ & $-24 \pm 4$ & $89 \pm 4$ & $1.2 \pm 0.1$ & $83 \pm 4$ & $107$ \\
2007.661 & $239 \pm 17$ & $3.2 \pm 0.2$ & $-30 \pm 4$ & $134 \pm 13$ & $1.0 \pm 0.1$ & $107 \pm 4$ & $137$ \\
2007.773 & $268 \pm 8$ & $4.5 \pm 0.1$ & $-49 \pm 4$ & $93 \pm 6$ & $0.9 \pm 0.1$ & $96 \pm 3$ & $145$ \\
2007.907 & $236 \pm 8$ & $5.0 \pm 0.1$ & $-37 \pm 3$ & $118 \pm 6$ & $0.7 \pm 0.1$ & $99 \pm 5$ & $136$ \\ 
2008.025 & $229 \pm 8$ & $4.2 \pm 0.2$ & $-23 \pm 4$ & $121 \pm 7$ & $0.8 \pm 0.2$ & $106 \pm 4$ & $129$ \\
2008.137 & $243 \pm 6$ & $4.3 \pm 0.2$ & $-25 \pm 5$ & $117 \pm 9$ & $0.7 \pm 0.1$ & $103 \pm 5$ & $128$ \\
\hline
\end{tabular}
\end{minipage}
\end{table*}

\begin{table*}
\begin{minipage}{\linewidth}
\caption{The results of the 8\,GHz VLBA observations.}\footnotetext{The columns are the same as in Table \ref{tab:evpa5}.}
\label{tab:evpa8}
\centering
\begin{tabular}{c|ccc|ccc|c}
\hline
\hline
Epoch & \multicolumn{3}{c}{Core} & \multicolumn{3}{c}{Jet} & \\
 & $I_\textrm{c}$ [mJy] & $P_\textrm{c}$ [mJy] & $\chi_\textrm{c} [^\circ]$ & $I_\textrm{j}$ [mJy] & $P_\textrm{j}$ [mJy] & $\chi_\textrm{j} [^\circ]$ & $\vert \chi_\textrm{c}-\chi_\textrm{j} \vert [^\circ]$ \\
\hline
2007.540 & $285 \pm 10$ & $2.0 \pm 0.2$ & $12 \pm 17$ & $75 \pm 21$ & $1.7 \pm 0.1$ & $87 \pm 17$ & $75$ \\
2007.661 & $264 \pm 10$ & $3.8 \pm 0.4$ & $-23 \pm 21$ & $81 \pm 8$ & $1.2 \pm 0.2$ & $96 \pm 21$ & $119$ \\
2007.773 & $252 \pm 4$ & $4.8 \pm 0.1$ & $-56 \pm 4$ & $93 \pm 6$ & $1.2 \pm 0.2$ & $80 \pm 4$ & $136$ \\
2007.907 & $264 \pm 10$ & $4.1 \pm 0.2$ & $-41 \pm 3$ & $84 \pm 6$ & $1.0 \pm 0.2$ & $82 \pm 15$ & $123$ \\
2008.025& $256 \pm 3$ & $4.7 \pm 0.3$ & $-20 \pm 4$ & $89 \pm 5$ & $1.1 \pm 0.1$ & $83 \pm 3$ & $103$ \\
2008.137 & $264 \pm 10$ & $4.5 \pm 0.1$ & $-9 \pm 4$ & $75 \pm 10$ & $1.0 \pm 0.1$ & $92 \pm 3$ & $101$ \\
\hline
\end{tabular}
\end{minipage}
\end{table*}

\begin{table*}
\begin{minipage}{\linewidth}
\caption{The results of the 15\,GHz VLBA observations.}\footnotetext{The columns are the same as in Table \ref{tab:evpa5}.}
\label{tab:evpa15}
\centering
\begin{tabular}{c|ccc|ccc|c}
\hline
\hline
 Epoch & \multicolumn{3}{c}{Core} & \multicolumn{3}{c}{Jet} & \\
& $I_\textrm{c}$ [mJy] & $P_\textrm{c}$ [mJy] & $\chi_\textrm{c} [^\circ]$ & $I_\textrm{j}$ [mJy] & $P_\textrm{j}$ [mJy] & $\chi_\textrm{j} [^\circ]$ & $\vert \chi_\textrm{c}-\chi_\textrm{j} \vert [^\circ]$ \\
\hline
2007.540 & $257 \pm 2$ & $1.3 \pm 0.3$ & $75 \pm 10$ & $39 \pm 4$ & $0.5 \pm 0.3 $ & $96 \pm 10$ & $24$ \\
2007.661 & $250 \pm 20$ & $2.7 \pm 0.1$ & $-11 \pm 6$ & $35 \pm 4$ & - & - & - \\ 
2007.773 & $238 \pm 20$ & $3.0 \pm 0.1$ & $-48 \pm 5$ & $26 \pm 4$ & - & - & - \\ 
2007.907 & $261 \pm 11$ & $2.3 \pm 0.1$ & $-42 \pm 10$ & $33 \pm 3$ & $0.6 \pm 0.3$ & $64 \pm 10$ & $106$ \\ 
2008.025 & $267 \pm 10$ & $3.5 \pm 0.1$ & $-4 \pm 10$ & $36 \pm 2$ & $0.5 \pm 0.1$ & $113 \pm 10$ & $114$ \\
2008.137 & $251 \pm 3$ & $2.9 \pm 0.1$ & $2 \pm 5$ & $31 \pm 2$ & - & - & - \\  
\hline
\end{tabular}
\end{minipage}
\end{table*}

In Fig. \ref{fig:polpa}, we show the difference between the core and jet EVPA. Note that this quantity is not affected 
by a possible calibration error of the EVPA absolute orientation. 
(EVPA miscalibration might occur in at least one epoch at 8\,GHz, see Sect. \ref{obs}). 
It is clear from Fig. \ref{fig:polpa}, that the relative position-angle difference between the electric vectors of the two components 
varies significantly from epoch to epoch (and thus on timescales of $\sim 6$ weeks) at all three frequencies. 
At 5\,GHz, the relative EVPA of the two features changes by $\sim 50^\circ$, then it rotates back by $20^\circ$. At 8\,GHz, 
the trend is similar $\sim 60^\circ$ rotation during the first three epochs, then the EVPA difference decreases by more than $30^\circ$ to around
$100^\circ$ in the last epoch. In the following, we discuss these simultaneous change in the EVPA of two physically distinct VLBI
components.

\subsection{Discussion of the EVPA changes} \label{poldisc}

\begin{figure}
\resizebox{\hsize}{!}{\includegraphics{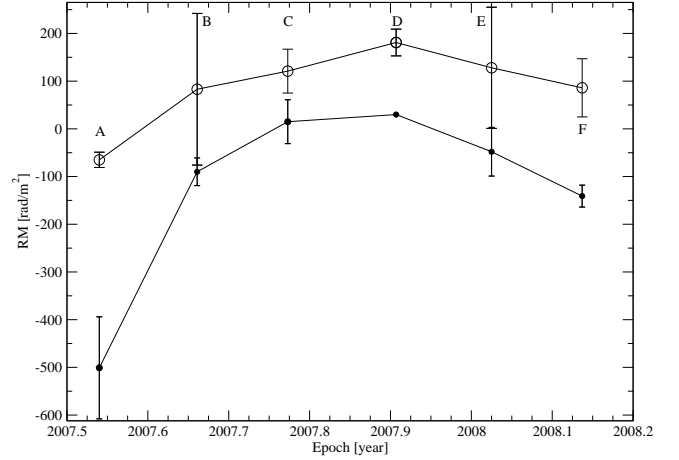}}
\caption{The rotation measure of the core (filled circles) and the jet (open circles) of J1128+592 during the VLBA observations. The VLBA epochs are denoted by letters (see Table \ref{tab:obs}).}
\label{fig:rm}
\end{figure}

\begin{figure}
\resizebox{\hsize}{!}{\includegraphics{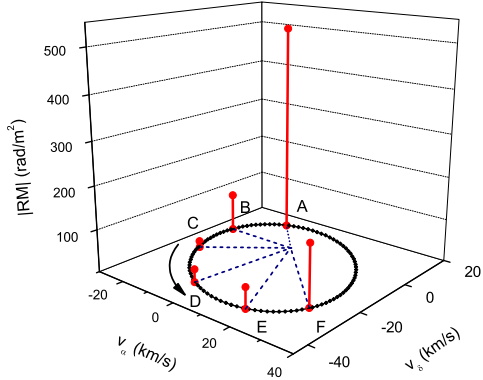}}
\caption{The variation in the rotation measure. The ellipse (solid black line) shows the apparent velocity of the screen relative to the Earth in the two orthogonal coordinates of right ascension and declination according to the fitted anisotropic annual modulation model. The dashed lines (blue) represent the velocity vectors at the times of the 6 VLBA observations (labeled A to F, see Table\ref{tab:obs}). The vertical lines in the plane orthogonal to the orbital motion indicate the absolute value of the RM at the times of the VLBA observations.}
\label{fig:vel}
\end{figure}

\begin{table}
\begin{minipage}{\linewidth}
\renewcommand{\thefootnote}{\thempfootnote}
  \caption{RM of the core (Col. 3) and jet component (Col. 4).}
  \label{tab:rm}
  \centering
  \begin{tabular}{cc|cc}
    \hline
    \hline
     & Epoch & $RM_\textrm{core}$ [rad/$\textrm{m}^2$] & $RM_\textrm{jet}$ [rad/$\textrm{m}^2$] \\
    \hline
    A & 2007.540 & $-501 \pm 107$ & $-65 \pm 16$\\ 
    B & 2007.661 & $-90 \pm 29$ & $(83 \pm 159)$\footnote{The values in parentheses are based upon only two frequency measurements.} \\ 
    C & 2007.773 & $15 \pm 46$ & $(121 \pm 46)$\footnotemark[\value{mpfootnote}] \\ 
    D & 2007.907 & $30 \pm 1$ & $181 \pm 28$ \\ 
    E & 2008.025 & $-48 \pm 51$ & $128 \pm 127$ \\ 
    F & 2008.137 & $-141 \pm 23$ & $(86 \pm 61)$\footnotemark[\value{mpfootnote}]  \\
    \hline
  \end{tabular}
\end{minipage}
\end{table}

In general, changes in the EVPA with frequency are explained by the rotation measure of an ionized plasma.
When polarized light passes through a magnetized plasma,
it is subject to Faraday rotation: the intrinsic polarization angle $\chi_{0}$ rotates by an amount proportional to the square of 
the observed wavelength \citep{faraday}. Thus, the observed polarization angle is $\chi=\chi_0+\textrm{RM} \lambda^2$.
The RM  depends on both the free electron density $n_e$ of the plasma along the line of sight and the line-of-sight component of 
the magnetic field $B_\parallel$: $\textrm{RM} \sim\int n_e B_\parallel ds$.

For the jet component, we detected polarization at 15\,GHz
in only three epochs, and we can therefore only obtain a reliable RM value for
these epochs. We note that plotting the EVPA versus lambda$^2$, all data points lie along a
straight line with relative little scatter and are therefore aligned
without requiring 180 degree ambiguity corrections.
We also estimated the RM for the other epochs with polarization detections at
only two frequencies. Naturally, these values are only approximate estimates of   
the RM (shown in parenthesis in Table \ref{tab:rm}).
However, it can be seen from Fig. \ref{fig:rm} that the RM changes smoothly from epoch to epoch.

The RM values obtained are displayed in Fig. \ref{fig:rm}. The RM of the core varied from 
$\sim -500 \textrm{\,rad/m}^2$ to $\sim+30 \textrm{\,rad/m}^2$, then gradually changed back to a negative value of $\sim-141 \textrm{\,rad/m}^2$.
The RM of the jet also exhibited variations, although less pronounced. Despite the relatively large measurement errors, it appears
as if the RM of the core and the jet vary in unison.

Temporal variability of the RM has been reported for only a few AGNs and on timescales of typically 
1 - 3 years: \object{3C273} \citep{asada_rm,zavala_rm}, 
\object{3C279} \citep[][and references therein]{zavala_rm} and \object{3C120} \citep{gomez_rm}.
The Faraday screen responsible for the rotation of the EVPA in these sources is understood to be 
source intrinsic, i.e., either circumnuclear (related to the narrow line region or the torus surrounding the AGN), or internal, 
meaning that it is physically associated with the jet \citep[e.g., related to a sheath around the jet,][]{inoue_rm}. 
According to \cite{gomez_rm} (and references therein), the internal Faraday rotation causes depolarization in the source.
Thus, when the rotation reaches an angle of $\sim 45^\circ$, the fractional polarization decreases by a factor of two. This effect is
not seen for J1128+592 in our data, therefore an external Faraday screen is more plausible. 

From Fig. \ref{fig:rm}, it is obvious that the apparent variation in the RM of core and jet
exhibit similar trends, despite their projected linear distance of $\geq 5$\,pc. 
This suggests that the core and jet are both covered by the same Faraday 
screen. The relative offsets between the RM of the jet and core, and 
the slightly more pronounced variability in the RM of the core component is indicative of
a second Faraday screen, which covers only the core region 
but not the jet. We tried to decompose the RM into two Faraday screens, one that
is common to both core and jet, and causes a similar variability trend throughout our six VLBA epochs 
($\Delta \textrm{RM} \sim 150 \textrm{\,rad/m}^2$ on a timescale of 5 to 6 months), 
and a second component, which only affects the core-region of the VLBI structure. This latter 
Faraday screen causes an RM variation in the core-region of 
$\Delta \textrm{RM}\leq 300 \textrm{\,rad/m}^2$ on shorter timescales of 1.5 months. 
At this point, it is obvious that this decomposition is affected by the relatively large 
measurement uncertainties and should be confirmed by future more finely (in time
and frequency) sampled measurements.

We may hypothesize that the screen affecting both components identically
could be the same local screen that is responsible for the IDV and the annual modulation. 
In this case, the observed systematic variation in
the RM could be caused by: (i) variations in the parallel component 
of the interstellar magnetic field $B_\parallel$, (ii) variation in the electron density $n_e$ in the local screen, or (iii) a
variation in the path length, i.e., the thickness of the screen with respect to its projection
on the ecliptic (orbital plane), or of course, a combination of these three effects.
If we attribute most of the observed systematic RM variability to a screen
in the nearby galactic ISM, changes in the magnetic field and/or the electron density 
over spatial scales of shorter than 2 AU (or timescales of a few weeks) would indicate
the presence of MHD-turbulence in the ISM. On the other hand, if the magnetic field is homogeneous 
on AU-scales and frozen in, it will not vary strongly (with neither time nor on AU-size scales).
The projection angle of the magnetic field ($B_\parallel \sim B_0 \sin(\phi)$), could however 
in principle vary if the screen is very close, because the Earth orbits
the Sun. For a presumed screen distance of $\sim 37$\,pc, one expects
variations in the projection angle of no more than $\sim \arcsin (2 \textrm{AU} / 37 \textrm{pc}) \sim 10^{-5~}~^\circ$,
which is far too small to cause any significant variation in the RM. 
It is also difficult to explain a fractional RM variation of $\Delta$ RM $\sim$ 150 rad/m$^2$
solely by changes in the path length. This points towards 
a combination of time-variable path length, electron density, and perhaps
even magnetic field, in such a way that it would explain the observed 
systematic RM changes, as illustrated by the 3D-diagram in Fig. \ref{fig:vel}.
We note that this time dependence would be the natural consequence of 
a refractive lens moving through the line of sight. These lenses are used to explain 
the so-called ``extreme scattering events'' in some IDV sources \citep{ese_model, ese_model2, ese_model3}
and cause systematic intensity variations on timescales of months.
\cite{ese_model} discuss the geometrical effects of such a lens and mention
the possibility of RM variations, which, in their case, were not however detected.
Since the frequency-dependent ``bending'' of the rays by the plasma lens
leads to caustics and  differential amplification of the (polarized) 
substructure of the source, it is possible that such a lens 
also introduces apparent RM variations. The discussion of this possibility,
however, is beyond the scope of this paper and deserves a future, more thorough elaboration.
The variation in the line-of-sight ``thickness''
of the screen (or the related emission measure, which is $\sim \int n_e ds$)
should also produce variation in the scattering measure (the line-of-sight integrated rms electron density) and the related
variability index. However, the limited accuracy of the variability-index
measurements illustrates these variations only very marginally (see solid circles in Fig. 3). 
According to \cite{local}, the RM of ISM clouds located inside the Local Bubble 
is typically of the order of $1 - 10$\,rad/m$^2$. This is much lower than the
observed RM variations towards J1128+592, or even the part of RM variations, which is the same
in both jet and core according to the decomposition analysis.

At this point, the signal-to-noise ratio of our data (i.e., RM, IDV timescale, and IDV amplitude) is clearly
insufficient to disentangle the RM variations caused locally (interstellar
weather, line-of-sight effects) and in the source itself. Higher polarization sensitivity,
a more accurate determination of the RM of VLBI core and jet, and a longer
time coverage (several orbital periods) will be necessary to study this effect further.

In an alternative and perhaps more conservative approach, one can argue that we do not see an intrinsically variable RM, 
nor does the intrinsic EVPA of the VLBI components change with time: we are observing instead at different times
blends of different subcomponents, which cannot be resolved in our VLBI observations, even at 15\,GHz. 
This blending of polarized components were reported in the case of other IDV sources, e.g., PKS0405-385 \citep{acf_def}.
Thus, the EVPA changes in J1128+592 
might also be explained by the emergence of a new polarized feature. As the new polarized subcomponent gradually separates from the core,
the core EVPA changes, since it is dominated more and more by the new, strongly polarized feature. As this feature expands and thus fades, 
the measured EVPA becomes dominated again by the original core polarization, and thus it rotates back towards its original value. 
This scenario would naturally explain the observed shift between the total intensity and polarized intensity features in the first 
epoch at 15\,GHz. \citep[Similar offsets are often reported in quasars and blazars, and explained as polarized subcomponents e.g.][]{bach_vsop,pol_subcomp}. However, this scenario would not explain the EVPA variations (and the similar trend in the core and jet RM variations) of the jet component. 
Moreover, we also observed a variation in the RM of the core and jet without 
significant time shift. This would be difficult to understand, 
if the RM of the core were affected by blending effects, e.g., because of 
a newly emerging polarized jet component.

A solution to this problem and a viable interpretation of the correlated variability in the EVPAs 
and the RM in core and jet, must most likely include two effects:
(i) a Faraday screen, which covers the entire VLBI structure and leads to
the observed similar variability trend in the RM of jet and core, and (ii) another process, which only affects the frequency
alignment of the EVPAs in the VLBI core region, and for which only marginal evidence
is seen in our data. At present, it is unclear whether this second process is
related to blending effects in a time-variable polarized substructure of the core
or to another Faraday screen (in the quasar), which affects only the core but not the jet
(circumnuclear screen, e.g., a torus). Future polarization sensitive VLBI monitoring of higher sensitivity, denser frequency coverage, 
and higher angular resolution -- i.e., as provided by the future space VLBI \citep[VSOP2, e.g.][]{vsop2} --
should help us to distinguish between source-extrinsic and source-intrinsic depolarization effects.

\section{Summary \label{end}}

We have presented new results derived from our 3 frequency (5/8/15 GHz) polarization-sensitive VLBA observations of the IDV source J1128+592. An aim of these observations was to discover whether source structural changes are responsible for the decrease in the strength of IDV observed 
in this source during 2007-2008. 

Our data uncovered an east-west oriented jet-core structure of the source. For the entire six epochs of VLBA data (spanning 0.6 yrs), 
we were unable to detect any significant change in the total intensity VLBI structure of the source. Compared with archival VLBA data, we found that the total flux density of the core changed significantly. This agrees with our single-dish monitoring result, which included a flare-like brightening in J1128+592 during early 2006. Compared to archival data, the size of the core at 5\,GHz also increased significantly, which may explain the quenching of
the intraday variability. Although we did not detect any structural changes in the source during our observations, the core size was found to be excessively large at 15\,GHz. This may be indicative of the emergence of a new jet component that is still blended with the core at the time of our observation. 

We fitted an annual modulation model to the IDV timescales derived from our densely time-sampled IDV flux-density monitoring data. One important parameter of the annual modulation model is the scattering lengthscale, which is determined by the scintillating source size and distance to scattering screen. From the IDV monitoring and the size limits of the VLBI core, we obtained a lower limit to the distance of the scattering screen of $\sim 37$\,pc. 

The source structure detected by our VLBA observations has an orientation that roughly agrees with the orientation 
of the anisotropic scattering obtained from the fit of annual modulation model. This might indicate that the source-intrinsic structure is responsible for the anisotropic scattering \citep[unlike in other well-known IDV sources, e.g.,][and references therein]{aniso}.

We found significant variations in the orientation of the polarization vector in the source (core and jet) with time. 
At all three observing frequencies, the EVPA and the deduced RM of the core varied in unison on timescales as short as 6 weeks. 
The variation in the RM of the jet is less pronounced than in the core 
and affected by larger measurement errors, but seems to follow the same systematic
trend as the RM variability of the core. The relatively fast variation of the RM and the similar trend
in RM variations of the core and jet might indicate that at least a certain fraction of the RM variability 
is caused by a nearby ISM screen, which can also be responsible for the IDV and
the annual modulation of the variability timescales. 
At this point, other source-intrinsic interpretations of the RM variability (e.g., internal Faraday rotation,
different RM in different regions of the source) cannot be excluded.
More data with superior time and frequency coverage will be necessary to distinguish between these possibilities.
  
\begin{acknowledgements}
The National Radio Astronomy Observatory is a facility of the National Science Foundation operated under cooperative agreement by Associated Universities Inc. This paper made use of data obtained with the 100\,m Effelsberg radio telescope of the Max-Planck-Institut f\"ur Radioastronomie (Bonn, Germany)
and the 25\,m Urumqi Observatory (UO) of the National Astronomical Observatories (NAOC) of the Chinese Academy of Sciences (CAS).
This research is made using data of the University of Michigan Radio Astronomy Observatory which is supported by funds of the University of Michigan. This work has benefited from research funding from the Hungarian Scientific Research Fund (OTKA, grant No. K72515). K.\'E.G. acknowledges fellowship received from the Japan Society for the Promotion of Science. N.M. has been supported for this research through a stipend from the International Max-Planck Research School (IMPRS) for Radio and Infrared Astronomy at the Universities of Bonn and Cologne. 
\end{acknowledgements}

\bibliographystyle{aa}
\bibliography{ref}

\begin{thebibliography}{39}
\expandafter\ifx\csname natexlab\endcsname\relax\def\natexlab#1{#1}\fi

\bibitem[{{Aller} {et~al.}(2003){Aller}, {Aller}, {Latimer}, \&
  {Hughes}}]{umrao_ref}
{Aller}, H.~D., {Aller}, M.~F., {Latimer}, G.~E., \& {Hughes}, P.~A. 2003, in
  Bulletin of the American Astronomical Society, Vol.~35, Bulletin of the
  American Astronomical Society, 723--+

\bibitem[{{Asada} {et~al.}(2008){Asada}, {Inoue}, {Kameno}, \&
  {Nagai}}]{asada_rm}
{Asada}, K., {Inoue}, M., {Kameno}, S., \& {Nagai}, H. 2008, \apj, 675, 79

\bibitem[{{Bach} {et~al.}(2006){Bach}, {Krichbaum}, {Kraus}, {Witzel}, \&
  {Zensus}}]{bach_vsop}
{Bach}, U., {Krichbaum}, T.~P., {Kraus}, A., {Witzel}, A., \& {Zensus}, J.~A.
  2006, \aap, 452, 83

\bibitem[{{Beckert} {et~al.}(2002){Beckert}, {Fuhrmann}, {Cim{\`o}},
  {Krichbaum}, {Witzel}, \& {Zensus}}]{beckert_evn}
{Beckert}, T., {Fuhrmann}, L., {Cim{\`o}}, G., {et~al.} 2002, in Proceedings of
  the 6th EVN Symposium, ed. E.~{Ros}, R.~W. {Porcas}, A.~P. {Lobanov}, \&
  J.~A. {Zensus}, 79

\bibitem[{{Bernhart} {et~al.}(2006){Bernhart}, {Krichbaum}, {Fuhrmann}, \&
  {Kraus}}]{cease_0917_4}
{Bernhart}, S., {Krichbaum}, T.~P., {Fuhrmann}, L., \& {Kraus}, A. 2006, ArXiv
  Astrophysics e-prints

\bibitem[{{Bignall} {et~al.}(2006){Bignall}, {Macquart}, {Jauncey}, {Lovell},
  {Tzioumis}, \& {Kedziora-Chudczer}}]{bignall_newest}
{Bignall}, H.~E., {Macquart}, J.~., {Jauncey}, D.~L., {et~al.} 2006, \apj, 652,
  1050

\bibitem[{{Blandford} \& {K{\"o}nigl}(1979)}]{jet_model}
{Blandford}, R.~D. \& {K{\"o}nigl}, A. 1979, \apj, 232, 34

\bibitem[{{Burn}(1966)}]{faraday}
{Burn}, B.~J. 1966, \mnras, 133, 67

\bibitem[{{Carter} {et~al.}(2009){Carter}, {Ellingsen}, {Macquart}, \&
  {Lovell}}]{new_ann}
{Carter}, S.~J.~B., {Ellingsen}, S.~P., {Macquart}, J.-P., \& {Lovell},
  J.~E.~J. 2009, ArXiv e-prints

\bibitem[{{Clegg} {et~al.}(1998){Clegg}, {Fey}, \& {Lazio}}]{ese_model}
{Clegg}, A.~W., {Fey}, A.~L., \& {Lazio}, T.~J.~W. 1998, \apj, 496, 253

\bibitem[{{Dennett-Thorpe} \& {de Bruyn}(2003)}]{annual1819}
{Dennett-Thorpe}, J. \& {de Bruyn}, A.~G. 2003, \aap, 404, 113

\bibitem[{{Fuhrmann} {et~al.}(2002){Fuhrmann}, {Krichbaum}, {Cim{\`o}},
  {Beckert}, {Kraus}, {Witzel}, {Zensus}, {QIAN}, \& {Rickett}}]{cease_0917}
{Fuhrmann}, L., {Krichbaum}, T.~P., {Cim{\`o}}, G., {et~al.} 2002, Publications
  of the Astronomical Society of Australia, 19, 64

\bibitem[{{Gab{\'a}nyi} {et~al.}(2007{\natexlab{a}}){Gab{\'a}nyi}, {Marchili},
  {Krichbaum}, {Britzen}, {Fuhrmann}, {Witzel}, {Zensus}, {M{\"u}ller}, {Liu},
  {Song}, {Han}, \& {Sun}}]{1128_AN}
{Gab{\'a}nyi}, K.~{\'E}., {Marchili}, N., {Krichbaum}, T.~P., {et~al.}
  2007{\natexlab{a}}, Astronomische Nachrichten, 328, 863

\bibitem[{{Gab{\'a}nyi} {et~al.}(2007{\natexlab{b}}){Gab{\'a}nyi}, {Marchili},
  {Krichbaum}, {Britzen}, {Fuhrmann}, {Witzel}, {Zensus}, {M{\"u}ller}, {Liu},
  {Song}, {Han}, \& {Sun}}]{1128_aa}
{Gab{\'a}nyi}, K.~{\'E}., {Marchili}, N., {Krichbaum}, T.~P., {et~al.}
  2007{\natexlab{b}}, \aap, 470, 83

\bibitem[{{Gab{\'a}nyi} {et~al.}(2008){Gab{\'a}nyi}, {Marchili}, {Krichbaum},
  {Fuhrmann}, {Britzen}, {Witzel}, {Zensus}, {Liu}, {Song}, {Han}, \&
  {Sun}}]{1128_kerastari}
{Gab{\'a}nyi}, K.~{\'E}., {Marchili}, N., {Krichbaum}, T.~P., {et~al.} 2008, in
  Proceedings of the Workshop Bursts, Pulses and Flickering: wide-field
  monitoring of the dynamic radio sky, accepted, ed. T.~{Tzioumis} \&
  J.~{Lazio}

\bibitem[{{Gabuzda} \& {G{\'o}mez}(2001)}]{pol_subcomp}
{Gabuzda}, D.~C. \& {G{\'o}mez}, J.~L. 2001, \mnras, 320, L49

\bibitem[{{G{\'o}mez} {et~al.}(2008){G{\'o}mez}, {Marscher}, {Jorstad},
  {Agudo}, \& {Roca-Sogorb}}]{gomez_rm}
{G{\'o}mez}, J.~L., {Marscher}, A.~P., {Jorstad}, S.~G., {Agudo}, I., \&
  {Roca-Sogorb}, M. 2008, \apjl, 681, L69

\bibitem[{{Heeschen} {et~al.}(1987){Heeschen}, {Krichbaum}, {Schalinski}, \&
  {Witzel}}]{idv_discovery2}
{Heeschen}, D.~S., {Krichbaum}, T., {Schalinski}, C.~J., \& {Witzel}, A. 1987,
  \aj, 94, 1493

\bibitem[{Hummel(1987)}]{var_VLBI}
Hummel, C.~A. 1987, Diploma thesis, Rheinische Friedrich-Wilhelms-Universit\"at
  Bonn

\bibitem[{{Inoue} {et~al.}(2003){Inoue}, {Asada}, \& {Uchida}}]{inoue_rm}
{Inoue}, M., {Asada}, K., \& {Uchida}, Y. 2003, in Astronomical Society of the
  Pacific Conference Series, Vol. 300, Radio Astronomy at the Fringe, ed. J.~A.
  {Zensus}, M.~H. {Cohen}, \& E.~{Ros}, 141--+

\bibitem[{{Jauncey} {et~al.}(2003){Jauncey}, {Johnston}, {Bignall}, {Lovell},
  {Kedziora-Chudczer}, {Tzioumis}, \& {Macquart}}]{annual1519_1}
{Jauncey}, D.~L., {Johnston}, H.~M., {Bignall}, H.~E., {et~al.} 2003, \apss,
  288, 63

\bibitem[{{Jauncey} \& {Macquart}(2001)}]{0917annual2}
{Jauncey}, D.~L. \& {Macquart}, J.-P. 2001, \aap, 370, L9

\bibitem[{{Kedziora-Chudczer}(2006)}]{0405annual}
{Kedziora-Chudczer}, L. 2006, \mnras, 369, 449

\bibitem[{Kellermann \& Pauliny-Toth(1969)}]{compton}
Kellermann, K.~I. \& Pauliny-Toth, I. I.~K. 1969, \apj, 155, L71

\bibitem[{{Kraus} {et~al.}(1999){Kraus}, {Witzel}, {Krichbaum}, {Lobanov},
  {Peng}, \& {Ros}}]{cease_0917_2}
{Kraus}, A., {Witzel}, A., {Krichbaum}, T.~P., {et~al.} 1999, \aap, 352, L107

\bibitem[{{Krichbaum} {et~al.}(2002){Krichbaum}, {Kraus}, {Fuhrmann},
  {Cim{\`o}}, \& {Witzel}}]{cease_0917_3}
{Krichbaum}, T.~P., {Kraus}, A., {Fuhrmann}, L., {Cim{\`o}}, G., \& {Witzel},
  A. 2002, Publications of the Astronomical Society of Australia, 19, 14

\bibitem[{{Lazio} {et~al.}(2000){Lazio}, {Fey}, {Dennison}, {Mantovani},
  {Simonetti}, {Alberdi}, {Foley}, {Fiedler}, {Garrett}, {Hirabayashi},
  {Jauncey}, {Johnston}, {Marcaide}, {Migenes}, {Nicolson}, \&
  {Venturi}}]{ese_model2}
{Lazio}, T.~J.~W., {Fey}, A.~L., {Dennison}, B., {et~al.} 2000, \apj, 534, 706

\bibitem[{{Rickett} {et~al.}(2006){Rickett}, {Lazio}, \& {Ghigo}}]{GBI}
{Rickett}, B., {Lazio}, T.~J.~W., \& {Ghigo}, F.~D. 2006, \apjs, 165, 439

\bibitem[{{Rickett} {et~al.}(2002){Rickett}, {Kedziora-Chudczer}, \&
  {Jauncey}}]{acf_def}
{Rickett}, B.~J., {Kedziora-Chudczer}, L., \& {Jauncey}, D.~L. 2002, \apj, 581,
  103

\bibitem[{{Rickett} {et~al.}(1995){Rickett}, {Quirrenbach}, {Wegner},
  {Krichbaum}, \& {Witzel}}]{acf_def2}
{Rickett}, B.~J., {Quirrenbach}, A., {Wegner}, R., {Krichbaum}, T.~P., \&
  {Witzel}, A. 1995, \aap, 293, 479

\bibitem[{{Rickett} {et~al.}(2001){Rickett}, {Witzel}, {Kraus}, {Krichbaum}, \&
  {Qian}}]{0917annual1}
{Rickett}, B.~J., {Witzel}, A., {Kraus}, A., {Krichbaum}, T.~P., \& {Qian},
  S.~J. 2001, \apjl, 550, L11

\bibitem[{{Senkbeil} {et~al.}(2008){Senkbeil}, {Ellingsen}, {Lovell},
  {Macquart}, {Cim{\`o}}, \& {Jauncey}}]{ese_model3}
{Senkbeil}, C.~E., {Ellingsen}, S.~P., {Lovell}, J.~E.~J., {et~al.} 2008,
  \apjl, 672, L95

\bibitem[{{Spangler}(2009)}]{local}
{Spangler}, S.~R. 2009, Space Science Reviews, 143, 277

\bibitem[{{Tsuboi}(2008)}]{vsop2}
{Tsuboi}, M. 2008, Journal of Physics Conference Series, 131, 012048

\bibitem[{{Walker} {et~al.}(2009){Walker}, {de Bruyn}, \& {Bignall}}]{aniso}
{Walker}, M., {de Bruyn}, G., \& {Bignall}, H. 2009, \mnras, 397, 447

\bibitem[{{Witzel} {et~al.}(1986){Witzel}, {Heeschen}, {Schalinski}, \&
  {Krichbaum}}]{idv_discovery}
{Witzel}, A., {Heeschen}, D.~S., {Schalinski}, C., \& {Krichbaum}, T. 1986,
  Mitteilungen der Astronomischen Gesellschaft Hamburg, 65, 239

\bibitem[{{Wu} {et~al.}(2008){Wu}, {Zhou}, {Ma}, {Wu}, {Jiang}, \&
  {Chen}}]{utolag1}
{Wu}, J., {Zhou}, X., {Ma}, J., {et~al.} 2008, \aj, 135, 258

\bibitem[{{Wu} {et~al.}(2009){Wu}, {Zhou}, {Ma}, {Wu}, {Zhang}, {Jiang}, \&
  {Chen}}]{utolag2}
{Wu}, J., {Zhou}, X., {Ma}, J., {et~al.} 2009, \aj, 137, 3961

\bibitem[{{Zavala} \& {Taylor}(2001)}]{zavala_rm}
{Zavala}, R.~T. \& {Taylor}, G.~B. 2001, \apjl, 550, L147

\end{thebibliography}

\end{document}